\documentclass{emulateapj}
\usepackage{natbib}
\usepackage{apjfonts}

\newcommand{\nobs}{$182$}
\newcommand{\goodnobs}{$38$}
\newcommand{\nred}{$18$}

\newcommand{\medsigv}{$1.3$}
\newcommand{\medsigteff}{$464$}
\newcommand{\medsiglogg}{$0.54$}
\newcommand{\medsigfeh}{$0.45$}
\newcommand{\members}{$18$}
\newcommand{\minsigv}{$0.4$}
\newcommand{\maxsigv}{$3.3$}
\newcommand{\minsigteff}{$50$}
\newcommand{\maxsigteff}{$1035$}
\newcommand{\minsiglogg}{$0.08$}
\newcommand{\maxsiglogg}{$1.30$}
\newcommand{\minsigfeh}{$0.04$}
\newcommand{\maxsigfeh}{$1.05$}

\newcommand{\twilightvoffset}{$-3.9$}
\newcommand{\twilightteffoffset}{$  69$}
\newcommand{\twilightloggoffset}{$ 0.09$}
\newcommand{\twilightfehoffset}{$-0.20$}
\newcommand{\twilightvscatter}{$ 0.1$}
\newcommand{\twilightteffscatter}{$  43$}
\newcommand{\twilightloggscatter}{$ 0.04$}
\newcommand{\twilightfehscatter}{$ 0.02$}

\newcommand{\vmean}{$64.8_{-1.0}^{+1.1}$}
\newcommand{\vdisp}{$3.6_{-0.6}^{+0.9}$}
\newcommand{\vgrad}{$0.4_{-0.2}^{+0.4}$}
\newcommand{\vtheta}{$-79_{-55}^{+220}$}
\newcommand{\fehmean}{$-2.67_{-0.34}^{+0.34}$}
\newcommand{\fehdisp}{$0.50_{-0.13}^{+0.17}$}
\newcommand{\fehgrad}{$0.02_{-0.06}^{+0.06}$}
\newcommand{\vmeanexpanded}{$64.8_{-1.0(-2.0)}^{+1.1(+2.2)}$}
\newcommand{\vdispexpanded}{$3.6_{-0.6(-1.1)}^{+0.9(+2.1)}$}
\newcommand{\vgradexpanded}{$0.4_{-0.2(-0.3)}^{+0.4(+1.0)}$}
\newcommand{\vthetaexpanded}{$-79_{-55(-93)}^{+220(+251)}$}
\newcommand{\fehmeanexpanded}{$-2.67_{-0.34(-0.67)}^{+0.34(+0.72)}$}
\newcommand{\fehdispexpanded}{$0.50_{-0.13(-0.23)}^{+0.17(+0.40)}$}
\newcommand{\fehgradexpanded}{$0.02_{-0.06(-0.13)}^{+0.06(+0.12)}$}

\newcommand{\doublestarv}{$120.4\pm0.7$}

\newcommand{\doublestarlogg}{$ 4.52\pm 0.23$}
\newcommand{\doublestarfeh}{$-0.52\pm 0.10$}
\newcommand{\doublestarvtwo}{$ 54.8\pm0.6$}

\newcommand{\doublestarloggtwo}{$ 4.41\pm 0.21$}
\newcommand{\doublestarfehtwo}{$-0.37\pm 0.08$}

\newcommand{\mrhalf}{$2.4_{-0.8}^{+1.3}\times 10^5$}
\newcommand{\mlratio}{$462_{-157}^{+264}$}

\newcommand{\nsubexposures}{$   73$}
\newcommand{\nsubexposuresredspectra}{$   34$}
\newcommand{\nsubexposuresbluespectra}{$   39$}

\newcommand{\bvoffset}{$  2.1\pm 0.3$}
\newcommand{\rvoffset}{$  1.1\pm 0.3$}

\newcommand{\teff}{T_{\mathrm{eff}}}
\newcommand{\feh}{\mathrm{[Fe/H]}}

\newcommand{\logg}{\log g}

\newcommand{\los}{\mathrm{los}}

\usepackage{amsmath}

\begin{document}
\title{Magellan/M2FS Spectroscopy of the Reticulum 2 Dwarf Spheroidal Galaxy\footnote{This paper presents data gathered with the Magellan Telescopes at Las Campanas Observatory, Chile.}}
\shorttitle{Magellan/M2FS Spectroscopy of Reticulum 2}
\author{Matthew G. Walker\altaffilmark{1}, Mario Mateo\altaffilmark{2}, Edward W. Olszewski\altaffilmark{3}, John I. Bailey III\altaffilmark{2}, Sergey E. Koposov\altaffilmark{4}, Vasily Belokurov\altaffilmark{4} and N.  Wyn Evans\altaffilmark{4}}
\email{mgwalker@andrew.cmu.edu}
\altaffiltext{1}{McWilliams Center for Cosmology, Department of Physics, Carnegie Mellon University, 5000 Forbes Ave., Pittsburgh, PA 15213, United States}
\altaffiltext{2}{Department of Astronomy, University of Michigan, 311 West Hall, 1085 S. University Ave., Ann Arbor, MI 48109}
\altaffiltext{3}{Steward Observatory, The University of Arizona, 933 N. Cherry Ave., Tucson, AZ 85721}
\altaffiltext{4}{Institute of Astronomy, University of Cambridge, Madingley Road, Cambridge, CB3 0HA, United Kingdom}

\begin{abstract} 
We present results from spectroscopic observations with the Michigan/Magellan Fiber System (M2FS) of \nobs\ stellar targets along the line of sight to the newly-discovered `ultrafaint' object Reticulum 2 (Ret 2).  For \goodnobs\ of these targets, the spectra are sufficient to provide simultaneous estimates of line-of-sight velocity ($v_{\los}$, median random error $\delta_{v_{\los}}=$ \medsigv\ km s$^{-1}$), effective temperature ($\teff$, $\delta_{\teff}=$ \medsigteff\ K), surface gravity ($\logg$, $\delta_{\logg}=$ \medsiglogg\ dex) and iron abundance ($\feh$, $\delta_{\feh}=$ \medsigfeh\ dex).  We use these results to confirm \members\ stars as members of Ret 2.  From the member sample we estimate a velocity dispersion of $\sigma_{v_{\los}}=$ \vdisp\ km s$^{-1}$ about a mean of $\langle v_{\los}\rangle =$ \vmean\ km s$^{-1}$ in the solar rest frame ($\sim -90.9$ km s$^{-1}$ in the Galactic rest frame), and a metallicity dispersion of $\sigma_{\feh}=$ \fehdisp\ dex about a mean of $\langle \feh \rangle =$ \fehmean.  These estimates marginalize over possible velocity and metallicity gradients, which are consistent with zero.  Our results place Ret 2 on chemodynamical scaling relations followed by the Milky Way's dwarf-galactic satellites.  Under assumptions of dynamic equilibrium and negligible contamination from binary stars---both of which must be checked with deeper imaging and repeat spectroscopic observations---the estimated velocity dispersion suggests a dynamical mass of $M(R_{\rm h})\approx 5R_{\rm h}\sigma_{v_{\los}}^2/(2G)=$ \mrhalf\ $M_{\odot}$ enclosed within projected halflight radius $R_{\rm h}\sim 32$ pc, with mass-to-light ratio $\approx 2M(R_{\rm h})/L_{\rm V}=$ \mlratio\ in solar units. 
\end{abstract}


\section{Introduction}

The census of Local Group galaxies has been revised dramatically and repeatedly over the last decade.  Mining of the SDSS stellar catalog has yielded discoveries of $\sim 15$ low-luminosity, dwarf-galactic satellites of the Milky Way \citep[e.g.,][]{willman05b,zucker06a,belokurov07}.  The PanDAS and PanStarrs surveys have found nearly two dozen new satellites of M31 \citep{mcconnachie09,martin13}.  All told, the population of known Local Group galaxies has nearly tripled \citep{mcconnachie12}.  Of the new members, perhaps the most intriguing are the Milky Way's `ultrafaint' satellites.  These objects have lowered the floor of the observed galaxy luminosity function from $M_V\sim -8$ to $M_V\sim -2$, such that some galaxies' total luminosities can be dominated by a single red giant star \citep{martin08}.  Moreover, the structural parameters of ultrafaints have blurred what was once an obvious distinction between the Milky Way's dwarf-galactic satellites and its globular clusters.  As a result, the proper classification of most ultrafaint objects now requires spectroscopic measurements of velocity dispersions, metallicities and metallicity dispersions that can indicate the presence of a dark matter halo.  

\smallskip
Most recently, the Dark Energy Survey (DES) has revealed nine new Galactic satellites at southern latitudes \citep[][`K15' and `DES15' hereafter]{koposov15,des15}.  Seven of the new objects have sizes and luminosities characteristic of ultrafaints.  One of them, Reticulum 2 (Ret 2), has already attracted attention for several reasons.  First, Ret 2 is the nearest ($D\sim 30$ kpc) and most easily detected of the newly-discovered objects.  Second, Ret 2 clearly has a flattened morphology, which may indicate ongoing tidal disruption or perhaps rotation.  Third, using public data from the Fermi-LAT, \citet{alex15} find evidence for gamma-ray emission that is consistent with dark matter annihilation in Ret 2.  \citet{fermi15} assign low significance to the gamma-ray signal based on unreleased Fermi-LAT data; however, in an independent analysis of the public data, \citet{hooper15} reproduce the original detection.  In any case, Ret 2 is clearly an intriguing object, and the first question to settle is whether Ret 2 presents chemo-dynamical evidence for dark matter.  That is, is Ret 2 a globular cluster or a galaxy?

\smallskip
Here we present results from an initial spectroscopic `reconnaissance' of individual stellar targets along the line of sight to Ret 2.  We identify a sample consisting of \nred\ member stars, which we use to characterize Ret 2's chemodynamical properties.  Specifically, we estimate the means and dispersions of velocity and metallicity distributions, and we check for velocity and metallicity gradients that might provide clues about Ret 2's dynamical state and formation history.  Finally, we compare Ret 2's properties to those of known dwarf galaxies and globular clusters in order to determine which population can claim Ret 2 as its newest member.

\section{Observations and Data Reduction}
\label{sec:obsreduce}
On 19 February 2015, we observed Ret 2 with the Michigan/Magellan Fiber System \citep[M2FS,][]{mateo12} on the 6.5-m Magellan/Clay Telescope at Las Campanas Observatory, Chile.  M2FS deploys up to 256 optical fibers---each with an entrance aperture of diameter $1.2$ arcsec---anywhere over a (f/11) focal surface with diameter $29$ arcmin.  The fibers feed twin spectrographs that can be configured independently in a wide variety of modes.  For the Ret 2 observations, both M2FS spectrographs were configured identically in `HiRes' mode, with order-isolation filters providing coverage over the range $5132-5186$ \AA\ at effective resolution $\mathcal{R}\sim 18000$.

\begin{figure*}
  \includegraphics[width=6.5in, trim=0in 2.5in 0.5in 0.5in,clip]{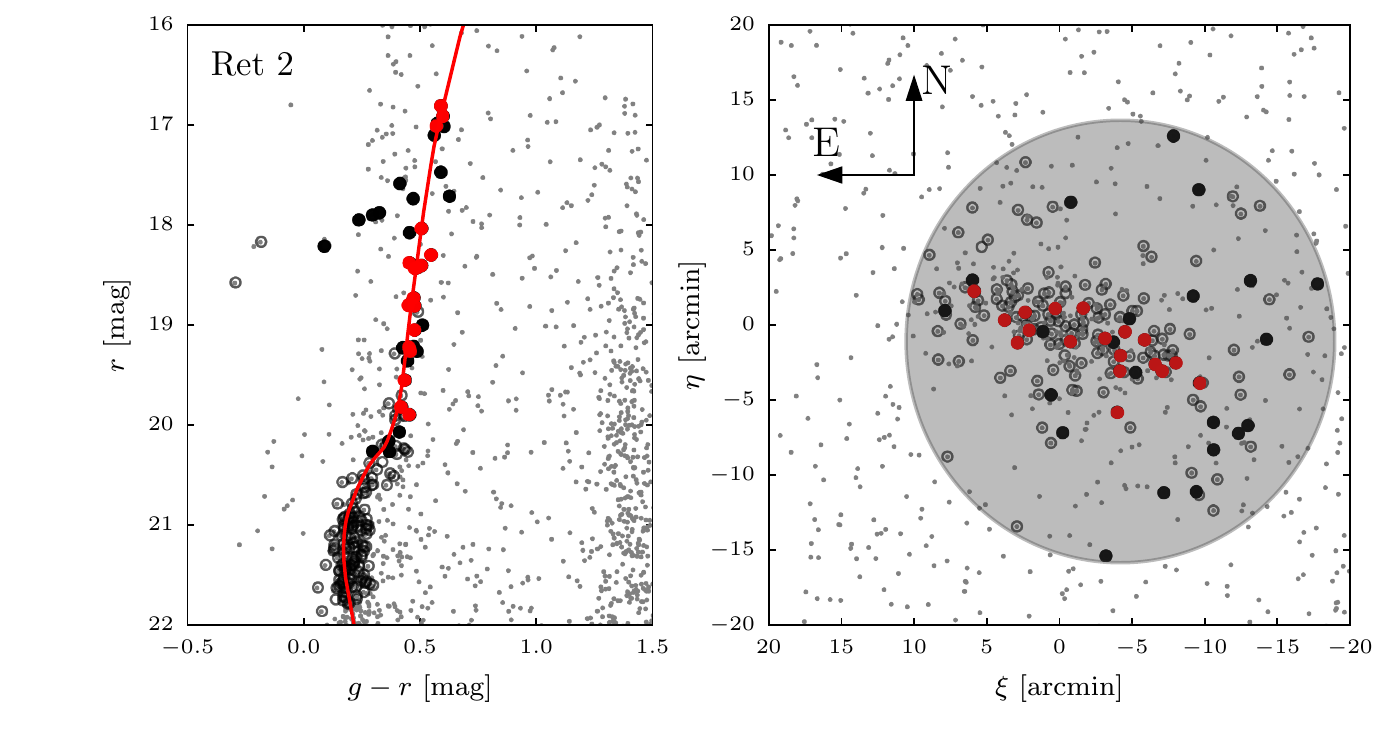}
  \caption{\textit{Left:} Color-magnitude diagram for stars within $R\leq 15$ arcmin of Ret 2's center \citep{koposov15}.  The red line is the Dartmouth isochrone for age $=12$ Gyr, [Fe/H]$=-2.5$, [$\alpha/$Fe]=+0.4, and $m-M=17.4$ \citep{dotter08}.  Large circles enclose stars identified as probable Ret 2 members by the DES MW working group (private communication, 2015) and subsequently targeted by M2FS.  Filled circles mark \goodnobs\ stars whose spectroscopic measurements meet quality-control criteria.  Of those, red circles identify stars whose spectroscopic properties are consistent with Ret 2 membership; black circles are spectroscopic non-members.   \textit{Right:} Standard coordinates for stars within $0.2$ magnitudes of the isochrone shown in the left-hand panel.  Markers and their colors have the same meanings as in the left-hand panel.  The large shaded circle represents the M2FS field of view.}
  \label{fig:ret2_map}
\end{figure*}

\smallskip
The left-hand panel of Figure \ref{fig:ret2_map} shows a color-magnitude diagram for stars projected within $15$ arcmin of Ret 2's center ($\alpha_{2000}=53.9256$ deg, $\delta_{2000}=-54.0492$ deg), from the photometric catalog derived by K15 in their analysis of public images from the Dark Energy Survey.  With M2FS we targeted \nobs\ stars (large circles in Figure \ref{fig:ret2_map}) selected as probable members of Ret 2 by the DES's Milky Way Science Working Group, who also provided the coordinates necessary for designing M2FS plug plates (private communication, 2015).  The right-hand panel of Figure \ref{fig:ret2_map} shows positions of the targeted stars.  We also allocated 64 fibers to regions of blank sky in order to estimate background.  We observed the Ret 2 field for a total of $2$ hours ($3\times 2400$ s).  At evening twilight the field was already rather low in the west, such that airmass during our observations increased from $1.3$ to $1.7$ and seeing hovered above $1.2$ arcsec.  

\smallskip
We analyze the individual images (i.e. the single 40-min exposures) as well as the `stacked' image consisting of the average over the three individual images.  Other raw images include ancillary exposures of emission-line sources (`arc' spectra), continuum sources (`quartz' spectra), twilight exposures, bias and dark exposures, all obtained using the same configuration and detector binning ($2\times 2$) as the science observations.  

\smallskip
We process the raw data using standard IRAF routines, following procedures described in detail by Mateo et al. (2015).  To summarize, we begin by performing overscan, bias and dark current corrections.  We then average the three individual exposures to produce a single, stacked science frame for each of the two spectrographs.  From science frames corresponding to individual as well as stacked exposures, we extract one-dimensional spectra.  For a given spectrum we obtain the wavelength solution by fitting a $4^{\rm th}$-order cubic spline to $\sim 30$ emission lines identified in the identically-extracted arc spectrum from the same aperture.  Residuals to the wavelength solution typically have rms $\sim 0.3$ km s$^{-1}$.  We correct for variations in fiber throughput by dividing each spectrum by a normalized fit to the continuum in the twilight spectrum obtained in the same aperture.  Finally, we estimate and subtract the mean contribution from sky background following the procedure introduced by \citet{koposov11}, which avoids re-binning science spectra.    

\smallskip
In parallel, we also compute variance spectra that account for all known sources of noise in the processed science spectra.  These include Poisson noise, readout noise, rms fluctuations in the bias and dark images, and background noise propagated from individual sky spectra into the mean sky spectrum.  We assign large variances ($10^{100}$) to pixels affected by cosmic rays, which we identify as outliers above an iterative fit to the continuum in the science spectrum.

\begin{figure*}
  \includegraphics[width=7in,trim=0.5in 1in 0.5in 3in,clip]{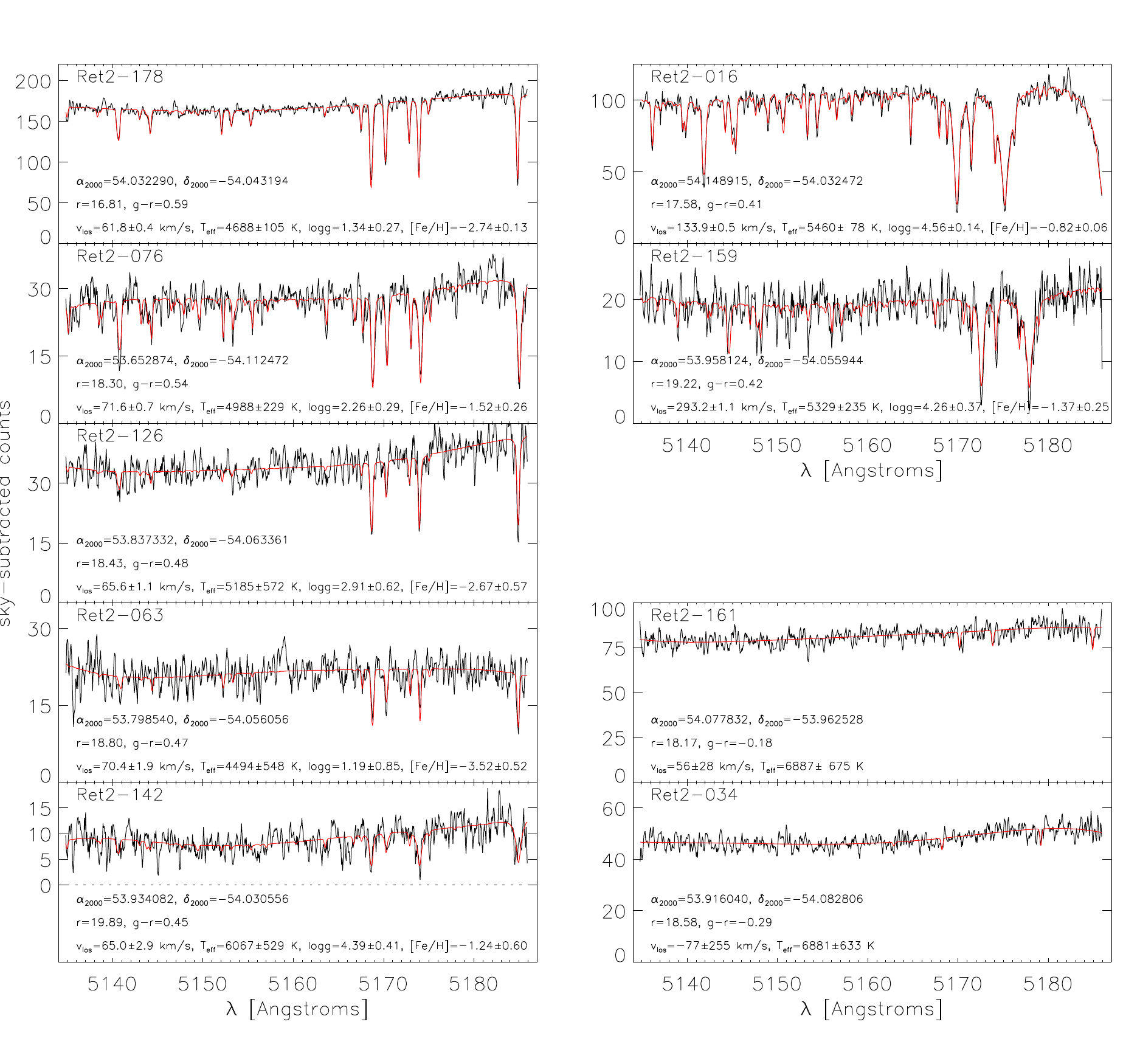}
  \caption{Sky-subtracted M2FS spectra, with best-fitting models overplotted.  Text gives target ID (upper left), equatorial coordinates (in degrees), color and magnitude, and estimates of redshift and stellar-atmospheric parameters.  Panels on the left-hand side show spectra for probable members of Reticulum 2.  The two upper-right panels show spectra for probable contaminants in the Galactic foreground.  Lower-right panels show spectra for two blue horizontal branch candidates.}
  \label{fig:ret2_specplot}
\end{figure*}
\begin{figure}
  \includegraphics[width=3.5in,trim=4in 1.2in 0.5in 5.25in,clip]{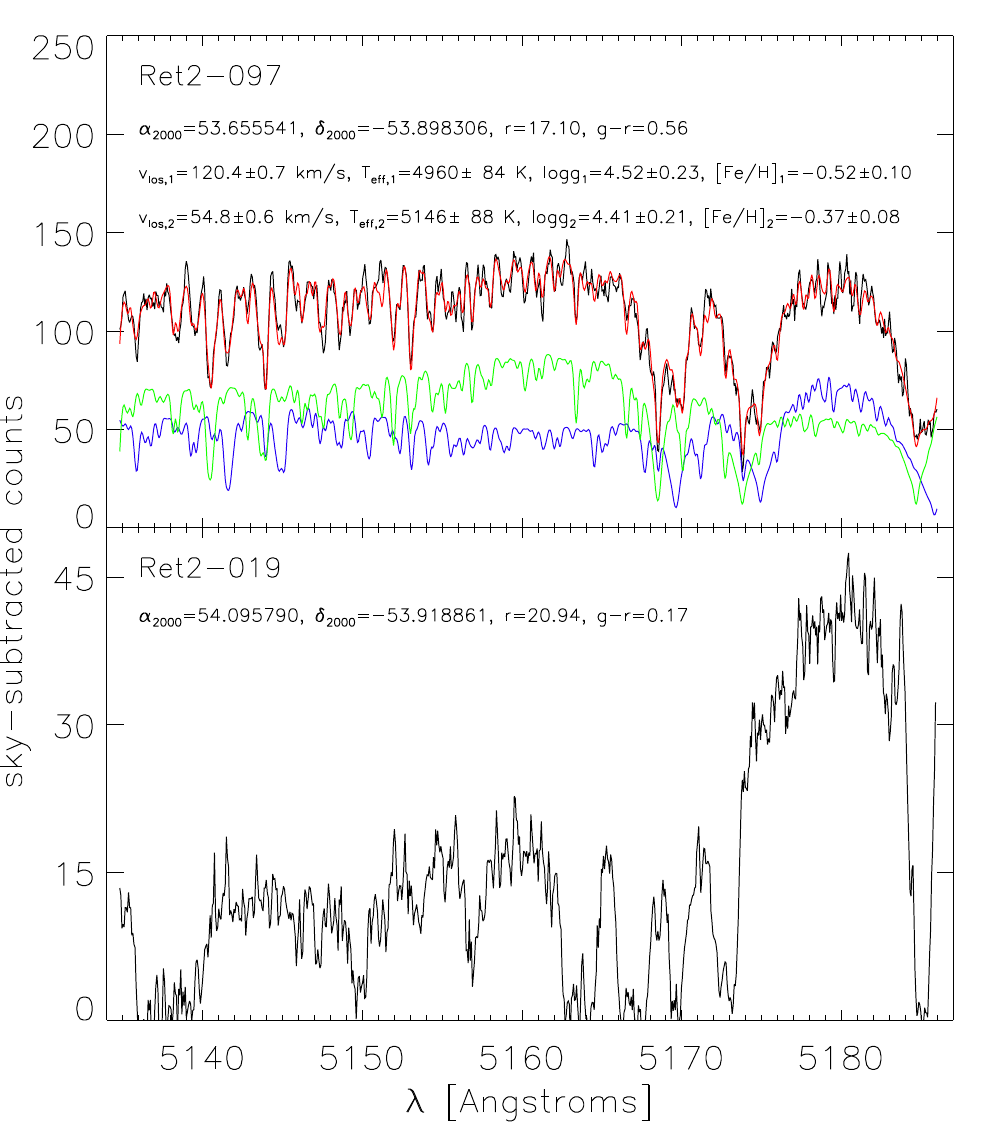}
  \caption{Sky-subtracted M2FS spectra for two anomalous sources.  \textit{Top:} This source is well-fit by the superposition (red) of two metal-rich model spectra (blue and green).  \textit{Bottom:} We speculate that the extremely wide, deep and irregular absorption features correspond to an absorption line system at redshift $z\ga 2.5$.}
  \label{fig:ret2_specialspecplot}
\end{figure}

\smallskip
Figure \ref{fig:ret2_specplot} displays examples of sky-subtracted M2FS spectra (from stacked science frames) for nine representative Ret 2 targets.  Those in left-hand panels are from likely members of Ret 2 (Section \ref{subsec:membership}) and show the relatively weak absorption features that are typical of metal-poor subgiants.  Spectra in upper right-hand panels belong to likely foreground contaminants and show the stronger absorption characteristic of late-type dwarfs.  Lower right-hand panels display spectra for two stars identified as blue horizontal branch candidates by both K15 and DES15.  These are the two bluest stars in our sample, and while their continua are well sampled, they exhibit no obvious absorption features in the observed spectral range (see Section \ref{subsec:specialcases}).  Finally, Figure \ref{fig:ret2_specialspecplot} shows two anomalous spectra that reveal their sources to be things other than single stars.  We discuss these spectra in more detail in Section \ref{subsec:specialcases}.

\section{Modelling of M2FS Spectra}
\label{sec:modelling}

We model the sky-subtracted M2FS
spectra following the procedure of \citet[][`W15' hereafter]{walker15}.  Briefly, we adopt a spectral model of
the form \citep[see also][]{koleva09,koposov11}
\begin{equation}
  M(\lambda)=P_l(\lambda)T\biggl (\lambda\biggl [1+\frac{Q_m(\lambda)+v_{\los}}{c}\biggr ]\biggr ),
  \label{eq:model}
\end{equation}
where $c$ is the speed of light, the polynomial $P_l(\lambda)$ gives shape to the continuum and $T\bigl (\lambda\bigl [1+\frac{Q_m(\lambda)+v_{\los}}{c}\bigr ]\bigr )$ is a continuum-normalized template spectrum.  The template is redshifted according to line-of-sight velocity, $v_{\rm los}$, and also by a  second polynomial, $Q_m(\lambda)$, that adjusts for systematic differences
between wavelength solutions of real and template spectra.  We generate templates from a synthetic library that has previously been used
for the SEGUE Stellar Parameter Pipeline \citep[`SSPP',][]{lee08a,lee08b} and which is calculated over a regular grid in effective temperature ($\teff$), surface gravity ($\logg$) and metallicity ($\feh$).  

\smallskip
Including nuisance parameters that specify polynomial coefficients and smooth the templates according to the instrumental line-spread function (LSF), the full model has 15 free parameters.  Four have physical meaning: $v_{\los}$, $\teff$, $\logg$ and $\feh$.  In order to obtain simultaneous estimates of all parameters, we follow W15's Bayesian analysis.  We adopt the same likelihood function given by W15's equation 9, and the same priors listed in W15's Table 2, with one exception: for $h_0$, the parameter that specifies the amount by which the template is smoothed in order to mimic the instrumental LSF, we adopt a prior that is uniform between $0.001$ \AA\ and $0.1$ \AA\ (W15's prior on $h_0$ is uniform between $0.05$ \AA\ and $0.1$ \AA).  We find that this adjustment improves fits to the narrowest absorption features.


\subsection{Posterior PDFs}
\label{subsec:posteriors}

We use the software package MultiNest \citep{feroz08,feroz09} to scan the parameter space and generate random samplings of the 15-dimensional posterior PDF.  
For a given parameter, $X$, we then calculate moments of the marginalized, 1D posterior PDF as follows.  The first moment is the mean, $\overline{X}\equiv N^{-1}\sum_{i=1}^Nx_i$, which we take to be the central value.  The second moment is the variance, $\sigma_X^2\equiv (N-1)^{-1}\sum_{i=1}^N(x_i-\overline{X})^2$, which we take to be the square of the $1\sigma$ credibility interval.  The third moment is skewness, $S\equiv N^{-1}\sum_{i=1}^N[(x_i-\overline{X})/\sqrt{\sigma_X^2}]^3$, which equals zero for a symmetric distribution.  The fourth moment is kurtosis, $K\equiv N^{-1}\sum_{i=1}^N[(x_i-\overline{X})/\sqrt{\sigma^2_X}]^4$, which distinguishes Gaussian distributions ($K=3$) from `leptokurtic' ones with sharper peaks and fatter tails ($K>3$) and `platykurtic' ones with broader peaks and weaker tails ($K<3$).  

\begin{figure*}
  \includegraphics[width=7in,trim=0.in 2.75in 1.5in 0.5in,clip]{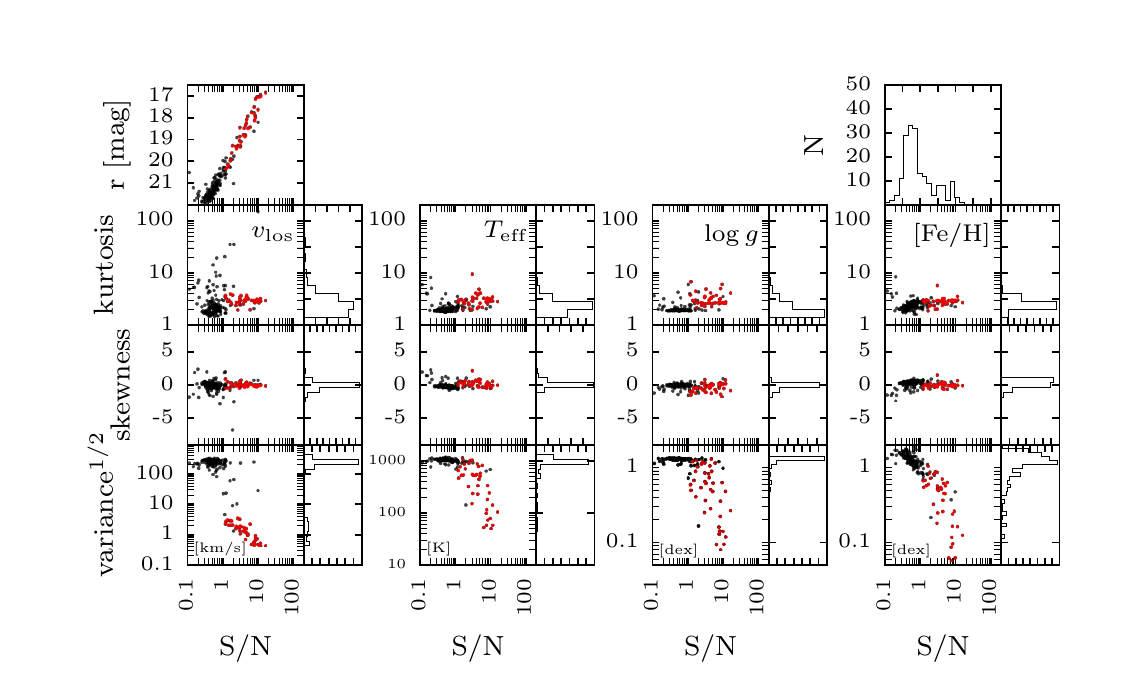}
  \caption{Moments of marginalized, 1D posterior PDFs of model parameters vs median S/N per pixel, for all \nobs\ spectra acquired with M2FS.  Histograms show marginalized distributions for each moment as well as S/N.  Red points indicate observations satisfying quality-control criteria (Section \ref{subsec:posteriors}), where posterior PDFs for velocity estimates are near-Gaussian ($|$skewness$|\leq 1$, $|$kurtosis$-3|\leq 1$) and the velocity error is $\sigma\leq 5$ km s$^{-1}$.}
  \label{fig:ret2_diagnostics}
\end{figure*}

For all \nobs\ stacked M2FS spectra from Ret 2 targets, Figure \ref{fig:ret2_diagnostics} shows how moments of PDFs for physical parameters vary with median S/N per pixel.  As S/N grows, the PDFs become narrower and more Gaussian ($S\sim 0$, $K\sim 3$).  In subsequent analysis we consider only the sample consisting of \goodnobs\ observations for which the PDF for velocity has $\sigma\leq 5$ km s$^{-1}$, $|S|\leq 1$ and $|K-3|\leq 1$ (red points in Figure \ref{fig:ret2_diagnostics}).  

\subsection{Accuracy and Precision}
\label{subsec:accuracy}
In order to examine accuracy and precision, we apply the same analysis to 256 solar spectra that we acquired during evening twilight on the same night and with the same spectrograph configuration used for the Ret 2 observations.  Reassuringly, parameter estimates from the collection of twilight spectra exhibit empirical scatter that is consistent with variances of the posterior PDFs.  Among the 256 twilight spectra, standard deviations of central values for $v_{\los}$, $\teff$, $\logg$ and $\feh$ are \twilightvscatter\ km s$^{-1}$, \twilightteffscatter\ K, \twilightloggscatter\ dex and \twilightfehscatter\ dex, respectively.  Furthermore, mean estimates show only mild systematic offsets with respect to known solar values: $\langle v_{\los}\rangle -v_{\los,\odot}=$ \twilightvoffset\  km s$^{-1}$, $\langle \teff \rangle-\teff$$_{\odot}=$ \twilightteffoffset\ K, $\langle \logg \rangle -\logg_{\odot}=$ \twilightloggoffset\ dex and $\langle \feh\rangle -\feh_{\odot}=$ \twilightfehoffset\ dex.  Following W15, for stellar-atmospheric parameters we treat these offsets as zero-point errors and subtract them from raw estimates obtained for science targets.  We also add the empirical scatter, in quadrature, to (square roots of) second moments for all observations.  After applying these adjustments, our estimates of stellar-atmospheric parameters have median (minimum, maximum) random errors of $\sigma_{\teff}=$ \medsigteff\ (\minsigteff, \maxsigteff) K, $\sigma_{\logg}=$ \medsiglogg\ (\minsiglogg, \maxsiglogg ) dex and $\sigma_{\feh}=$ \medsigfeh\ (\minsigfeh, \maxsigfeh ) dex.

\smallskip
For velocities, we do not adjust the zero point based on results from the twilight spectra.  In our experience, M2FS velocities shift systematically as ambient temperature changes rapidly during twilight, and the observed offset between twilight and solar velocities reflects this phenomenon.  As pointed out to us by the DES collaboration (private communication, 2015), a temperature-dependent velocity shift continued, albeit at a slower rate, during our observations of Ret 2.  In order to examine this effect, we compare velocity estimates that we obtain from fits to spectra obtained in individual science exposures.  We consider only the \nsubexposures\ single-exposure spectra whose results satisfy the same quality-control criteria imposed on the stacked spectra (Section \ref{subsec:posteriors}).  Consistently with the effect noticed by the DES collaboration, we find that velocity estimates shift systematically over the three exposures and at different rates for the two spectograph channels.  From \nsubexposuresbluespectra\ (\nsubexposuresredspectra) spectra observed  on M2FS's `blue' (`red') channel, we measure a weighted mean difference of $\langle v_{\rm exp3}-v_{\rm exp1}\rangle =$ \bvoffset km s$^{-1}$ (\rvoffset\ km s$^{-1}$) between the first and third exposures.  Thus the raw velocities that we estimate from the stacked frames have channel-dependent zero-point errors.   In order to compensate for this effect, we subtract the appropriate channel-specific value of $\langle v_{\rm exp3}-v_{\rm exp1}\rangle$ from each of the raw velocities that we estimate from the stacked frames.  We also add the corresponding error in our estimate of $\langle v_{\rm exp3}-v_{\rm exp1}\rangle$, in quadrature, to the error of our velocity estimates, after which our velocity sample has median (minimum, maximum) error $\sigma_{v_{\los}}=$ \medsigv\ (\minsigv, \maxsigv) km s$^{-1}$.  This procedure is designed only to remove the dependence of the velocity zero point on channel, and does not necessarily correct for a global error in zero point.  However, a comparison to an independent spectroscopic sample obtained with VLT/FLAMES (Koposov et al., in preparation), shows that the zero points are in agreement.  For the $17$ stars common to both samples, the mean velocity difference is $\langle v_{\rm M2FS}-v_{\rm VLT}\rangle =0.3\pm 0.3$ km s$^{-1}$.

\subsection{Repeatability}
\label{subsec:repeatability}
\begin{figure*}
  \includegraphics[width=6.5in, trim=0in 2.7in 0.5in 0.5in,clip]{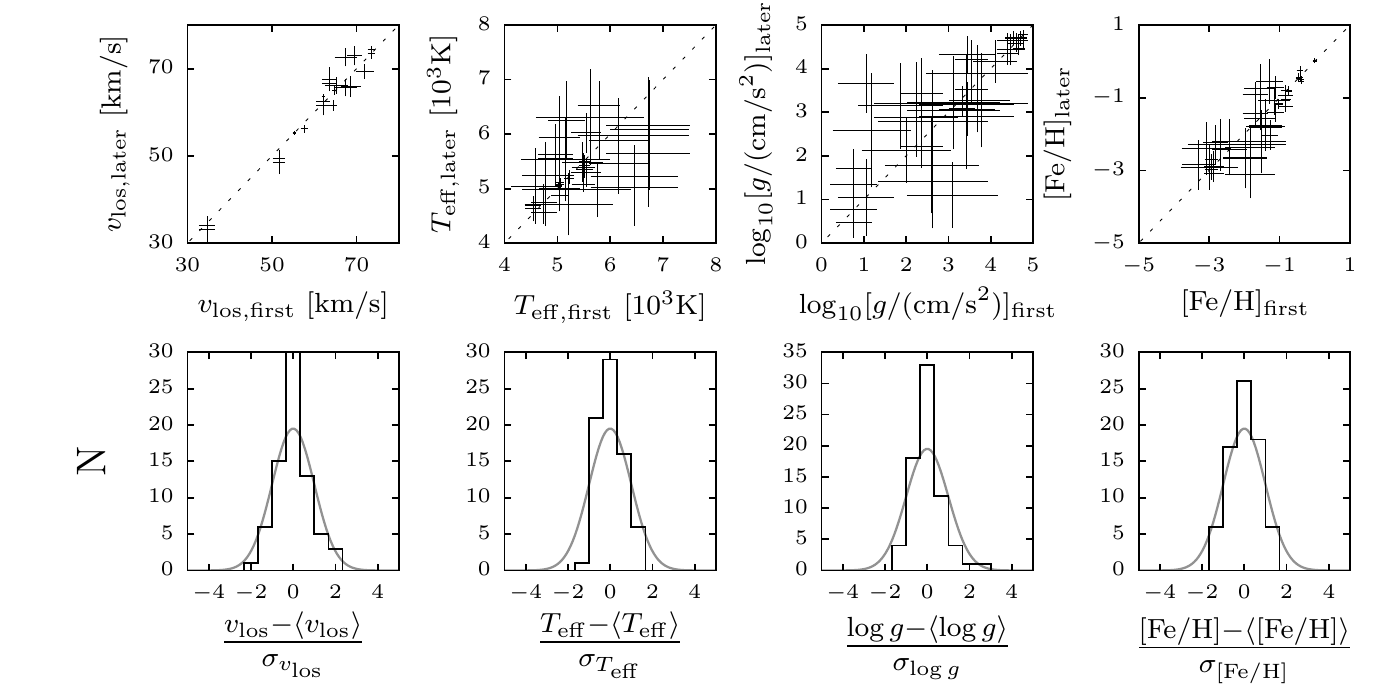}
  \caption{Repeatability of parameter estimates, from independent analyses of spectra extracted from three exposures.  Scatterplots in the top row display parameter estimates obtained from either of two later exposures against those obtained from the first exposure.  Dashed lines are 1:1 relations.  Histograms in the bottom row indicate distributions of deviations with respect to the weighted mean for each unique star, normalized by credibility intervals.  Overplotted are Gaussian distributions (gray curves) with integrated areas equal to those of the histograms.   The empirical distributions are generally narrower than the Gaussian ones, implying that our errors may be slightly overestimated.}
  \label{fig:ret2_repeats}
\end{figure*}

In order to gauge repeatability of our estimates for physical parameters, we again consider results from our modelling of \nsubexposures\ single-exposure spectra whose results satisfy quality-control criteria.  The top row of panels in Figure \ref{fig:ret2_repeats} compares parameter estimates from the first exposure with those in either of the two later exposures (all velocities from a given exposure and channel are shifted to remove the systematic drift in zero-point discussed above).  The bottom row of panels in Figure \ref{fig:ret2_repeats} displays distributions of deviations with repect to weighted means---$\langle \overline{X}\rangle \equiv \sum_{i=1}^N (\overline{X}_i/\sigma^2_{X_i})/\sum_{i=1}^{N_{\rm obs}}(1/\sigma_{X_i}^{2})$---normalized by propagated error.  Gray curves show Gaussian distributions with the same integrated area as the histograms.  The observed distributions are all narrower than the Gaussian curves, implying that the variances of the PDFs give credibility intervals that may be slightly overestimated.  

\smallskip
As a final check on reliability of our velocity estimates, we independently measure line-of-sight velocities using IRAF's `fxcor' package.  We cross-correlate each target spectrum from the stacked frame with a high-S/N M2FS spectrum that we acquired for the radial velocity standard star CPD-432527 ($v_{\los}=19.7\pm 0.9$ km s$^{-1}$; \citealt{udry99}) immediately before the Ret 2 observations.  Reassuringly, we find good agreement with results from the Bayesian procedure described above.  For the \goodnobs\ spectra that satisfy quality-control criteria (Section \ref{subsec:posteriors}), the mean deviation between (raw) velocity estimates is $\sim 0.1\pm 0.30$ km s$^{-1}$.  Furthermore, when we model the spectrum of CPD-432527 using our Bayesian procedure, we obtain $v_{\los}=19.7\pm 0.3$ km s$^{-1}$, recovering the previously-published value.

\subsection{Data}
\label{subsec:datatable}
Table \ref{tab:ret2_table1} lists results for the \goodnobs\ spectra that satisfy the quality-control criteria discussed in Section \ref{subsec:posteriors}.  The online version of this article includes a supplementary table with results for all \nobs\ observed stars.  The first three columns of Table \ref{tab:ret2_table1} list target ID and equatorial coordinates, followed by angular separation from Ret 2's center and $g$, $r$ magnitudes from the photometry of K15, corrected for extinction using the dust maps of \citet{schlegel98}.  The seventh column lists median S/N per pixel.  Columns 8-11 then list the results of our spectroscopic modelling in terms of the first four moments of PDFs for $v_{\los}$, $\teff$, $\logg$ and $\feh$.  For each parameter, the quoted central value is the first moment (mean) and the credible interval is the square root of the second moment; both values have been adjusted according to the offsets and empirical scatter described in Section \ref{subsec:accuracy}.  Skewness and kurtosis are listed parenthetically as superscripts above the credible interval.  The last column indicates whether or not the star satisfies membership criteria discussed in Section \ref{subsec:membership}.  For all observations, the first exposure began at heliocentric Julian date HJD=2457072.542 days and the midpoint of the $3\times 40$-minute exposure occured at HJD=2457072.579 days.

\begin{table*}
\scriptsize
\caption{M2FS Stellar Spectroscopy of Reticulum 2$^{a}$}
\begin{tabular}{@{}ccccccccccccccccccccccc@{}}
\hline
\\
ID&$\alpha_{2000}$&$\delta_{2000}$&$R$&$g$&$r$&S/N$^{b}$&$v_{\rm los}$&$T_{\rm eff}$&$\log_{10}g$&$\mathrm{[Fe/H]}$&member?\\

& [hh:mm:ss]&[$^{\circ}$:$\arcmin$:$\arcsec$]&[arcmin]&[mag]&[mag]&&[km s$^{-1}$]$^{c}$&[K]&[dex]$^{d}$&[dex]\\
\hline
Ret2-004&03:35:56.27&-54:03:16.2&$ 2.1$&$ 18.83$&$ 18.37$&$  5.7$&$  63.9\pm 2.3^{(  0.2,   2.0)}$&$6506\pm 811^{(-0.3, 2.3)}$&$3.12\pm1.14^{(-0.7, 2.7)}$&$-2.26\pm0.64^{(-0.6, 2.6)}$&Y\\
Ret2-006&03:36:01.76&-54:04:05.4&$ 3.1$&$ 19.52$&$ 19.05$&$  3.2$&$  62.5\pm 1.9^{(  0.7,   3.7)}$&$5313\pm 524^{( 0.4, 3.3)}$&$4.02\pm0.62^{(-1.1, 4.9)}$&$-2.88\pm0.50^{( 0.2, 3.0)}$&Y\\
Ret2-016&03:36:35.74&-54:01:56.9&$ 7.9$&$ 17.99$&$ 17.58$&$  9.5$&$ 133.9\pm 0.5^{(  0.0,   3.0)}$&$5460\pm  78^{( 0.0, 2.9)}$&$4.56\pm0.14^{( 0.1, 2.7)}$&$-0.82\pm0.06^{(-0.0, 3.1)}$&N\\
Ret2-021&03:36:22.83&-53:59:55.5&$ 6.7$&$ 18.53$&$ 18.07$&$  7.7$&$ 128.3\pm 0.5^{( -0.0,   2.7)}$&$5285\pm 115^{( 0.3, 3.2)}$&$3.19\pm0.15^{( 0.2, 3.1)}$&$-0.99\pm0.12^{( 0.2, 3.1)}$&N\\
Ret2-023&03:36:21.86&-54:00:40.6&$ 6.3$&$ 20.23$&$ 19.81$&$  1.6$&$  66.7\pm 2.0^{(  0.3,   4.0)}$&$5658\pm1034^{( 0.2, 2.0)}$&$3.37\pm0.89^{(-0.5, 3.1)}$&$-1.96\pm1.05^{(-0.1, 2.0)}$&Y\\
Ret2-032&03:35:40.69&-54:10:05.1&$ 7.1$&$ 18.19$&$ 17.90$&$  4.9$&$  49.0\pm 1.0^{( -0.2,   3.4)}$&$5786\pm 503^{( 0.3, 2.7)}$&$3.81\pm0.48^{( 0.0, 2.7)}$&$-1.75\pm0.45^{( 0.0, 2.6)}$&N\\
Ret2-035&03:35:37.06&-54:04:01.2&$ 1.3$&$ 18.54$&$ 18.03$&$  4.6$&$  62.9\pm 1.2^{(  0.4,   3.8)}$&$5034\pm 441^{( 0.9, 4.9)}$&$1.50\pm0.89^{( 0.2, 2.3)}$&$-2.90\pm0.45^{( 0.5, 3.3)}$&Y\\
Ret2-040&03:35:58.14&-54:02:04.7&$ 2.5$&$ 19.25$&$ 18.80$&$  4.1$&$  69.1\pm 1.5^{( -0.3,   3.1)}$&$6283\pm 865^{(-0.1, 2.2)}$&$2.97\pm1.05^{(-0.4, 2.7)}$&$-2.28\pm0.71^{(-0.3, 2.5)}$&Y\\
Ret2-049&03:35:20.36&-54:18:16.9&$15.7$&$ 20.55$&$ 20.26$&$  1.2$&$ 308.6\pm 2.3^{(  0.9,   3.7)}$&$6009\pm 645^{( 0.1, 2.8)}$&$4.06\pm0.59^{(-0.9, 3.8)}$&$-0.58\pm0.67^{(-0.4, 3.1)}$&N\\
Ret2-062&03:35:09.50&-54:02:29.6&$ 4.8$&$ 18.20$&$ 17.73$&$  7.3$&$  95.9\pm 0.6^{( -0.1,   3.0)}$&$5426\pm  98^{( 0.1, 2.8)}$&$4.49\pm0.17^{( 0.2, 2.8)}$&$-0.87\pm0.09^{( 0.0, 3.0)}$&N\\
Ret2-063&03:35:11.65&-54:03:21.8&$ 4.5$&$ 19.28$&$ 18.80$&$  3.0$&$  70.4\pm 1.9^{( -0.2,   3.5)}$&$4494\pm 548^{( 2.1, 9.2)}$&$1.19\pm0.85^{( 0.8, 3.5)}$&$-3.52\pm0.52^{( 1.5, 5.7)}$&Y\\
Ret2-069&03:35:02.49&-54:03:54.2&$ 5.9$&$ 19.20$&$ 18.72$&$  3.7$&$  67.9\pm 1.3^{( -0.0,   2.6)}$&$5048\pm 523^{( 0.8, 4.1)}$&$1.90\pm0.83^{(-0.1, 2.7)}$&$-2.77\pm0.55^{( 0.3, 2.9)}$&Y\\
Ret2-072&03:34:37.97&-54:13:59.6&$14.5$&$ 20.47$&$ 20.06$&$  1.4$&$ 137.3\pm 3.0^{(  0.5,   2.9)}$&$5574\pm 762^{( 0.6, 3.1)}$&$3.39\pm1.20^{(-0.9, 3.0)}$&$-2.13\pm0.68^{( 0.2, 2.7)}$&N\\
Ret2-076&03:34:36.69&-54:06:44.9&$10.3$&$ 18.84$&$ 18.30$&$  4.3$&$  71.6\pm 0.7^{(  0.0,   2.9)}$&$4988\pm 229^{( 0.7, 3.9)}$&$2.26\pm0.29^{( 0.1, 3.3)}$&$-1.52\pm0.26^{( 0.3, 3.3)}$&Y\\
Ret2-077&03:34:53.23&-54:14:03.9&$13.2$&$ 20.63$&$ 20.26$&$  1.2$&$ 290.8\pm 2.7^{( -0.4,   3.4)}$&$5372\pm 464^{( 0.3, 3.1)}$&$4.28\pm0.50^{(-1.4, 6.8)}$&$-1.57\pm0.55^{(-0.0, 2.8)}$&N\\
Ret2-079&03:34:54.24&-54:05:58.0&$ 7.6$&$ 18.91$&$ 18.42$&$  3.9$&$  70.0\pm 1.7^{( -0.2,   3.0)}$&$4915\pm 543^{( 0.7, 3.3)}$&$3.44\pm0.74^{(-0.3, 3.1)}$&$-3.19\pm0.52^{( 0.5, 2.9)}$&Y\\
Ret2-080&03:34:47.93&-54:05:25.0&$ 8.3$&$ 17.50$&$ 16.91$&$ 11.2$&$  63.2\pm 0.5^{(  0.1,   3.3)}$&$4501\pm 141^{( 0.6, 3.5)}$&$0.58\pm0.48^{( 0.7, 2.6)}$&$-3.16\pm0.16^{( 0.7, 3.6)}$&Y\\
Ret2-087&03:34:13.05&-53:59:56.1&$13.4$&$ 19.50$&$ 19.00$&$  2.9$&$ 200.5\pm 1.7^{( -0.4,   2.9)}$&$5211\pm 150^{( 0.0, 3.0)}$&$4.55\pm0.26^{(-0.8, 3.4)}$&$-1.32\pm0.18^{( 0.1, 3.0)}$&N\\
Ret2-089&03:34:05.48&-54:03:49.9&$14.2$&$ 17.61$&$ 17.01$&$ 10.1$&$  53.2\pm 0.5^{( -0.0,   2.7)}$&$5107\pm  50^{(-0.4, 3.1)}$&$4.77\pm0.10^{(-0.5, 2.7)}$&$-0.49\pm0.04^{( 0.0, 3.0)}$&N\\
Ret2-090&03:33:41.70&-54:00:07.2&$17.9$&$ 18.30$&$ 18.21$&$  4.5$&$  32.4\pm 1.6^{(  0.3,   3.5)}$&$6465\pm 773^{(-0.2, 2.3)}$&$3.43\pm1.30^{(-1.2, 3.4)}$&$-1.74\pm0.62^{(-0.5, 2.7)}$&N\\
Ret2-092&03:34:49.20&-53:50:19.7&$14.8$&$ 17.55$&$ 16.98$&$ 11.3$&$  16.2\pm 0.4^{(  0.0,   2.9)}$&$5218\pm  58^{( 0.0, 2.9)}$&$4.60\pm0.12^{( 0.2, 2.8)}$&$-0.01\pm0.05^{( 0.0, 3.1)}$&N\\
Ret2-100&03:34:39.81&-54:00:58.5&$ 9.4$&$ 18.18$&$ 17.95$&$  8.1$&$ 199.4\pm 0.9^{( -0.2,   2.9)}$&$6368\pm 333^{( 0.1, 2.6)}$&$4.05\pm0.36^{(-0.0, 2.7)}$&$-1.07\pm0.24^{(-0.1, 2.8)}$&N\\
Ret2-102&03:34:18.31&-54:10:06.2&$14.2$&$ 18.06$&$ 17.47$&$  7.5$&$  28.1\pm 0.5^{( -0.1,   3.0)}$&$5118\pm  58^{(-0.4, 3.0)}$&$4.70\pm0.13^{(-0.4, 2.6)}$&$-0.54\pm0.06^{(-0.0, 3.0)}$&N\\
Ret2-103&03:34:13.93&-54:09:34.4&$14.5$&$ 18.33$&$ 17.71$&$  6.3$&$  20.4\pm 0.5^{(  0.1,   3.0)}$&$5069\pm  52^{(-0.3, 3.3)}$&$4.79\pm0.10^{(-1.0, 3.8)}$&$-0.47\pm0.06^{(-0.0, 3.0)}$&N\\
Ret2-110&03:34:29.93&-54:11:11.7&$13.4$&$ 19.80$&$ 19.35$&$  2.4$&$ 290.5\pm 1.9^{( -0.1,   2.8)}$&$5797\pm 888^{( 0.5, 2.5)}$&$2.12\pm1.11^{( 0.3, 2.4)}$&$-1.69\pm0.82^{( 0.1, 2.4)}$&N\\
Ret2-111&03:34:30.06&-54:09:22.2&$12.4$&$ 18.20$&$ 17.88$&$  8.0$&$ 270.1\pm 0.7^{( -0.1,   2.9)}$&$5523\pm 182^{( 0.5, 3.4)}$&$3.63\pm0.23^{( 0.3, 3.2)}$&$-1.22\pm0.17^{( 0.2, 3.1)}$&N\\
Ret2-115&03:35:31.13&-54:01:48.2&$ 2.0$&$ 17.57$&$ 17.01$&$  9.0$&$  60.4\pm 0.7^{( -0.2,   3.2)}$&$5164\pm 242^{( 0.3, 3.3)}$&$2.44\pm0.63^{(-1.7, 6.0)}$&$-2.62\pm0.26^{( 0.1, 2.9)}$&Y\\
Ret2-126&03:35:20.96&-54:03:48.1&$ 3.2$&$ 18.91$&$ 18.43$&$  5.0$&$  65.6\pm 1.1^{( -0.0,   3.2)}$&$5185\pm 572^{( 0.8, 4.1)}$&$2.91\pm0.62^{( 0.0, 3.6)}$&$-2.67\pm0.57^{( 0.3, 2.9)}$&Y\\
Ret2-128&03:35:06.56&-54:06:04.3&$ 6.1$&$ 19.74$&$ 19.26$&$  2.3$&$ 160.5\pm 1.6^{(  0.3,   2.4)}$&$5466\pm 319^{( 0.2, 2.7)}$&$3.99\pm0.54^{(-0.5, 2.7)}$&$-1.02\pm0.33^{(-0.0, 2.9)}$&N\\
Ret2-129&03:34:57.57&-54:05:31.4&$ 7.0$&$ 18.90$&$ 18.40$&$  3.0$&$  61.8\pm 1.4^{(  0.3,   3.2)}$&$5296\pm 539^{( 0.5, 3.5)}$&$2.99\pm0.65^{(-0.1, 3.6)}$&$-2.04\pm0.57^{(-0.1, 2.9)}$&Y\\
Ret2-134&03:35:15.17&-54:08:43.0&$ 7.0$&$ 19.71$&$ 19.26$&$  3.0$&$  70.2\pm 3.3^{(  0.6,   2.5)}$&$5875\pm1019^{( 0.1, 2.0)}$&$1.94\pm1.15^{( 0.3, 2.3)}$&$-2.41\pm0.87^{(-0.1, 2.1)}$&Y\\
Ret2-136&03:35:14.01&-54:05:58.1&$ 5.1$&$ 19.98$&$ 19.55$&$  1.8$&$  66.4\pm 2.9^{(  0.3,   2.6)}$&$5582\pm 849^{( 0.4, 2.6)}$&$2.78\pm1.25^{(-0.3, 2.2)}$&$-2.14\pm0.84^{( 0.0, 2.4)}$&Y\\
Ret2-138&03:35:13.73&-54:04:56.7&$ 4.6$&$ 19.67$&$ 19.22$&$  1.9$&$  60.1\pm 2.1^{(  0.0,   3.8)}$&$5032\pm 694^{( 0.7, 3.1)}$&$2.97\pm1.02^{(-0.4, 2.7)}$&$-2.62\pm0.72^{( 0.3, 2.5)}$&Y\\
Ret2-142&03:35:44.18&-54:01:50.0&$ 1.2$&$ 20.34$&$ 19.89$&$  1.7$&$  65.0\pm 2.9^{( -0.2,   2.8)}$&$6067\pm 529^{(-0.1, 2.8)}$&$4.39\pm0.41^{(-1.2, 4.5)}$&$-1.24\pm0.60^{(-0.3, 3.0)}$&Y\\
Ret2-147&03:35:36.93&-53:54:45.1&$ 8.2$&$ 20.52$&$ 20.15$&$  1.5$&$ 302.3\pm 2.1^{( -0.5,   3.1)}$&$5335\pm 524^{( 0.5, 3.2)}$&$3.71\pm0.68^{(-0.4, 2.9)}$&$-1.23\pm0.58^{( 0.1, 3.1)}$&N\\
Ret2-153&03:35:46.16&-54:07:33.8&$ 4.6$&$ 19.21$&$ 18.74$&$  4.3$&$ 217.9\pm 1.2^{(  0.2,   3.1)}$&$5089\pm 333^{( 0.6, 4.0)}$&$2.82\pm0.51^{(-0.2, 4.2)}$&$-2.16\pm0.37^{( 0.1, 3.1)}$&N\\
Ret2-159&03:35:49.95&-54:03:21.4&$ 1.2$&$ 19.65$&$ 19.22$&$  3.1$&$ 293.2\pm 1.1^{(  0.2,   3.0)}$&$5329\pm 235^{( 0.2, 2.9)}$&$4.26\pm0.37^{(-0.4, 2.6)}$&$-1.37\pm0.25^{( 0.1, 2.9)}$&N\\
Ret2-178&03:36:07.75&-54:02:35.5&$ 3.8$&$ 17.39$&$ 16.81$&$ 15.6$&$  61.8\pm 0.4^{( -0.1,   3.0)}$&$4688\pm 105^{( 0.1, 2.9)}$&$1.34\pm0.27^{(-0.8, 4.2)}$&$-2.74\pm0.13^{(-0.0, 2.8)}$&Y\\
\hline
\end{tabular}
\\
\raggedright
$^{a}$This version lists results only for observations satisfying quality-control criteria (Section \ref{subsec:posteriors}).  The online version includes results for all targeted stars.\\
$^{b}$median signal-to-noise ratio per pixel\\
$^{c}$line-of-sight velocity in the heliocentric rest frame\\
$^{d}$units of $g$ are cm s$^{-2}$\\
\label{tab:ret2_table1}
\end{table*}

\subsection{Special Cases}
\label{subsec:specialcases}

Four spectra from our sample merit special attention.  Table \ref{tab:specialcases} lists their coordinates and $g$, $r$ magnitudes.
The first two (Ret2-161 and Ret2-034)  belong to the two blue horizontal branch (BHB) candidates and are displayed in the lower-right panels of Figure \ref{fig:ret2_specplot}.  Their spectra show well-sampled continua but no obvious absorption features, as expected given the greater degree of ionization at high temperature.  As a result, model parameters for these stars are poorly constrained.  While their velocities are loosely  consistent with Ret 2 membership (see Section \ref{subsec:membership}), the associated uncertainties are $\sim 30$ km s$^{-1}$ and $\sim 260$ km s$^{-1}$, such that neither spectrum satisfies our quality-control criteria (Section \ref{subsec:posteriors}).  On the other hand, their temperatures ($\teff$=$6890\pm 680$ K and $6880\pm 630$ K), while also highly uncertain, are the two largest that we obtained for any of our \nobs\ observed spectra.  We conclude that these two stars remain strong BHB candidates, with confirmation requiring spectra that cover a larger range in wavelength.

\smallskip
The next special case (Ret2-097) appears in the upper panel of Figure \ref{fig:ret2_specialspecplot}.  This `star' has color and magnitude placing it near the isochrone and above the horizontal branch in Figure \ref{fig:ret2_map}.  However, the spectrum is rich in broad and apparently double-valleyed absorption features.  We model this spectrum as the superposition of \textit{two} spectra of the form given by Equation \ref{eq:model}.  Allowing for a second set of stellar-atmospheric parameters and a second continuum polynomial, the resulting model has 25 free parameters.   As shown in Figure \ref{fig:ret2_specplot}, this double-star model gives a reasonably good fit to the spectrum.   One of the sources has $v_{\los}=$ \doublestarvtwo\ km s$^{-1}$, slightly below the range of velocities that we attribute to Ret 2 members (Section \ref{subsec:membership}).  Its strong surface gravity ($\logg=$ \doublestarloggtwo) and rich metallicity ($\feh=$ \doublestarfehtwo) are typical of foreground contamination.  The other contributing star has $v_{\los}=$ \doublestarv\ km s$^{-1}$, $\logg=$ \doublestarlogg\ and is also metal-rich, with $\feh=$ \doublestarfeh.  We speculate that this source is a physical binary composed of two K-type main sequence stars at a distance of a few kpc.  Assuming a mass ratio near unity, the center of mass has $v_{\los}\sim 100$ km s$^{-1}$, inconsistent with Ret 2 membership.  Due to the anomalous nature of this spectrum, we do not include these results in Table \ref{tab:ret2_table1} or consider them in subsequent analysis.  

\smallskip
Finally, the bottom panel of Figure \ref{fig:ret2_specialspecplot} displays a spectrum (Ret2-019) that is difficult to classify given the limited spectral range.  Here we consider a few possibilities.  First, the absorption features may have molecular origin, specifically from MgH, which has a bandhead at $\sim 5200$ \AA\ and extends blue-ward by $\sim 200$ \AA.  However, the features in our spectrum are too wide, too deep and too irregularly spaced to match the typical appearance of that band \citep{weck03}.  In addition, MgH absorption becomes evident typically in stars of spectral types K0 or later, much redder than the de-reddened color ($g-r\sim 0.17$) of this source.  Alternatively, this spectrum may reveal the presence of a complex absorption line system along the line of sight to a remote active galactic nucleus.  The line widths and extreme depths appear similar to numerous line complexes reported by \citet{srianand10} for various absorption lines in  Lyman-$\alpha$ systems.  Furthermore, the source's blue color is consistent with that of a background quasi-stellar object (QSO) at moderate redshift.  For a number of plausible UV lines that may produce the observed absorption features, the implied redshift of the absorber is $z\ga 2.5$, with the background source obviously more distant.  The nearest known extragalactic source in the NED catalog is detected in the UV, by GALEX, and is 24 arcsec away from our target.  A definitive classification of this source will require spectroscopy with broader wavelength coverage.
\begin{table}
  \begin{centering}
  \scriptsize
  \caption{Coordinates and photometry for special cases (Section \ref{subsec:specialcases})}
  \begin{tabular}{@{}lllllllll@{}}
    \hline
    ID&$\alpha_{2000}$&$\delta_{2000}$&$R$&$g$&$r$&notes\\
    & [hh:mm:ss]&[$^{\circ}$:$\arcmin$:$\arcsec$]&[arcmin]&[mag]&[mag]\\
    \hline
    Ret2-161&03:36:18.68&-53:57:45.1&$ 7.5$&$ 17.99$&$ 18.17$&BHB candidate\\
    Ret2-034&03:35:39.85&-54:04:58.1&$ 2.0$&$ 18.28$&$ 18.58$&BHB candidate\\
    Ret2-097&03:34:37.33&-53:53:53.9&$ 13.1$&$ 17.66$&$17.10 $&double star\\
    Ret2-019&03:36:22.99&-53:55:07.9&$ 9.9$&$ 21.11$&$ 20.94$&unknown\\
    \hline
    \\
  \end{tabular}
  \label{tab:specialcases}
  \end{centering}
\end{table}

\section{Results}
\label{sec:results}

Scatterplots in Figure \ref{fig:ret2_params} show how the spectroscopically-derived quantities ($v_{\los}$, $\teff$, $\logg$, $\feh$) relate to each other and to photometrically-derived quantities (position, color, magnitude).  

\subsection{Membership}
\label{subsec:membership}

\begin{figure*}
  \includegraphics[width=6.5in]{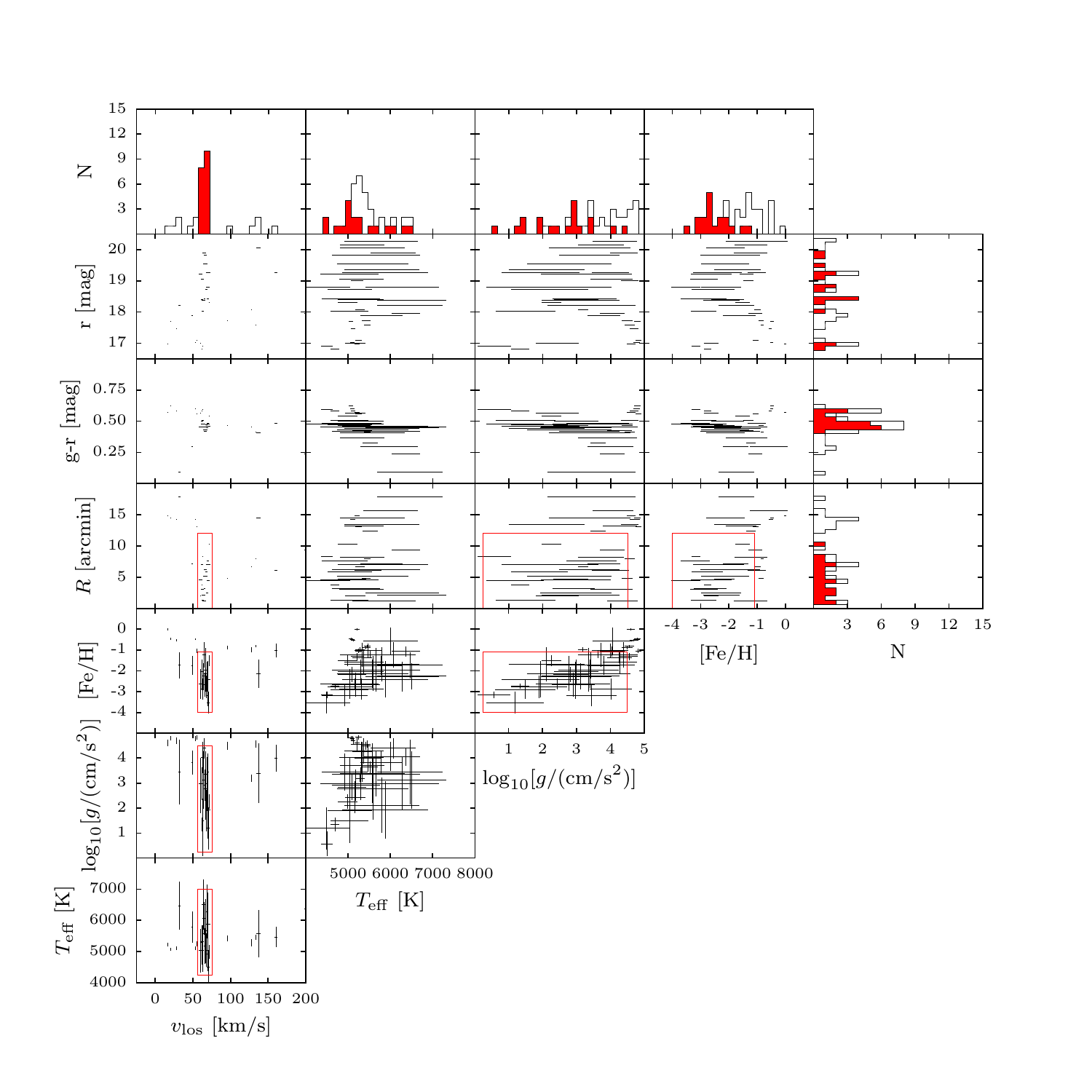}
  \caption{Scatterplots showing relations among photometrically- (magnitude, $r$, color, $g-r$, and separation, $R$, from Ret 2's center) and spectroscopically-derived ($v_{\los}$, $\teff$, $\logg$ and $\feh$) quantities for individual Ret 2 stars.  Red boxes are drawn, by eye, to enclose probable members.  For each observable, histograms show 1D distributions for the full sample (open black) and likely members (solid red).}
  \label{fig:ret2_params}
\end{figure*}

\smallskip
Figure \ref{fig:ret2_params} also helps to distinguish bona fide members of Ret 2 from contaminants in the Galactic foreground.  We expect Ret 2 members to exhibit a relatively narrow range in velocity, lower metallicities, lower surface gravities, and to be clustered nearer Ret 2's center than are their foreground counterparts.  Figure \ref{fig:ret2_params} clearly shows a population of stars having these characteristics.  Red boxes in Figure \ref{fig:ret2_params} enclose measurements that cluster near the centers of Ret 2's distributions for $v_{\los}$, $\logg$,  $\feh$ and position, as determined by eye.  We make no selection based on color, magnitude or temperature because the full ranges for these quantities are consistent with Ret 2 membership.  
We count \nred\ stars that are inside \textit{all} red boxes.  Table \ref{tab:ret2_table1} lists a membership status of `Y' for these stars, and `N' for stars that lie outside any of the red boxes.

\smallskip
In previous work we have employed an expectation-maximization (EM) algorithm in order to estimate model parameters in the presence of contamination \citep{walker09b}.  Under the assumption that velocity dispersion is spatially uniform, our EM algorithm provides estimates of membership probabilities for all individual stars.  Reassuringly, when we apply the EM algorithm to the Ret 2 data in Table \ref{tab:ret2_table1}, the sum of membership probabilities is 17.8, with each of the stars inside our red boxes receiving membership probability $>0.85$.  Furthermore, the EM algorithm's estimates of means and dispersions of velocity and metallicity distributions agree well with those obtained in the independent analysis that we present in Section \ref{subsec:chemodynamics}.  

\subsection{Chemodynamics of Reticulum 2}
\label{subsec:chemodynamics}

In order to characterize the internal chemodynamics of Ret 2, we consider a simple model in which the velocities and metallicities of the \members\ member stars are normally distributed about means that can vary systematically with position in order to account for velocity and metallicity gradients.  Specifically we assume that the mean line-of-sight velocity, $\mu_{v_{\los}}$, is a function of projected position ($R,\theta$),
\begin{equation}
  \mu_{v_{\los}}(R,\theta)\equiv \langle v_{\los}\rangle+k_{v_{\los}}R\cos(\theta-\theta_{v_{\los}}),
  \label{eq:meanvelocity}
\end{equation}
where $\langle v_{\los}\rangle$ is the mean velocity at the center, $k_{v_{\los}}\equiv |\mathrm{d}\mu_{v_{\los}}/\mathrm{d}R|$ is the magnitude of maximum velocity gradient and $\theta_{v_{\los}}$ (measured from north of Ret 2's center and opening to the east, in equatorial coordinates) specifies its direction.  We assume that mean metallicity, $\mu_{\feh}$, is a function only of separation from the center,
\begin{equation}
  \mu_{\feh}(R)\equiv \langle \feh\rangle+k_{\feh}R,
  \label{eq:meanfeh}
\end{equation}
where $\langle \feh\rangle $ is the mean metallicity at the center and $k_{\feh}\equiv \mathrm{d}\mu_{\feh}/\mathrm{d}R$ specifies the magnitude of any (isotropic) metallicity gradient.  

\smallskip
Under these assumptions, the joint probability density for observables $v_{\los}$ and $\feh$, for a star at position $(R,\theta)$, is
\begin{eqnarray}
  p\bigl (v_{\los},\feh \big | R,\theta\bigr )=\frac{(2\pi)^{-1}}{\sqrt{\bigl (\sigma^2_{v_\los}+\delta^2_{v_{\los}}\bigr )\bigl (\sigma^2_{\feh}+\delta^2_{\feh}\bigr )}} \nonumber\\   
  \times \exp\biggl [-\frac{1}{2}\frac{\bigl (v_{\los}-\mu_{v_{\los}} )^2}{\sigma^2_{v_\los}+\delta^2_{v_{\los}}} -\frac{1}{2}\frac{\bigl (\feh-\mu_{\feh}\bigr )^2}{\sigma^2_{\feh}+\delta^2_{\feh}}\biggr ]
\end{eqnarray}
where $\sigma_{v_{\los}}$ and $\sigma_{\feh}$ are velocity and metallicity dispersions, respectively, and $\delta_{v_{\los}}$ and $\delta_{\feh}$ are measurement errors.  Given a vector of free parameters $\vec{\theta}\equiv (\langle v_{\los}\rangle, \sigma_{v_{\los}}, k_{v_{\los}}, \theta_{v_{\los}}, \langle \feh\rangle ,\sigma_{\feh}, k_{\feh})$, a data set consisting of $N$ observations of Ret 2 members, $D\equiv \{(R_i,\theta_i,v_{\los,i},\feh_i) \}_{i=1}^N$, has likelihood
\begin{equation}
  \mathcal{L}(D\big |\vec{\theta})=\displaystyle\prod_{i=1}^Np(v_{\los,i},\feh_i\big |R_i,\theta_i).
\end{equation}
From Bayes' theorem, given the data, the model has posterior probability distribution function (PDF)
\begin{equation}
  p(\vec{\theta}\big |D)=\frac{\mathcal{L}(D\big |\vec{\theta})p(\vec{\theta})}{p(D)},
  \label{eq:bayes}
\end{equation}
where $p(\vec{\theta})$ is the prior PDF and $p(D)\equiv\int \mathcal{L}(D\big |\vec{\theta})p(\vec{\theta})d\vec{\theta}$ is the `evidence'.  

\begin{figure}
  \includegraphics[width=3.5in,trim=0.25in 0in 3.0in 1.1in,clip]{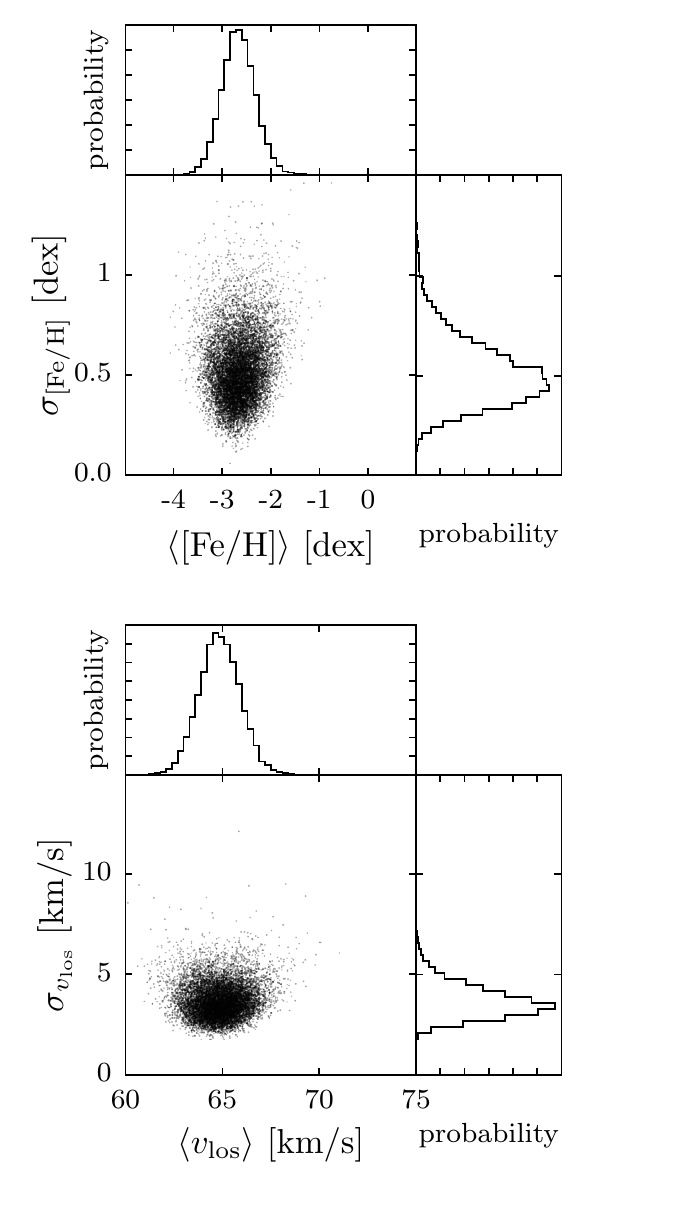}
  \caption{Samples drawn randomly from posterior PDFs for means and dispersions of Ret 2's metallicity (top) and velocity (bottom) distributions.  Histograms display marginalized, 1D PDFs for each parameter.}
  \label{fig:ret2_meansdispersions}
\end{figure}

\smallskip
Again we use MultiNest to scan the 7-dimensional parameter space and to draw random samples from the posterior PDF.  Figure \ref{fig:ret2_meansdispersions} displays random samplings from PDFs for the means and dispersions.  Table \ref{tab:gradient} summarizes the results.  For each of the seven free parameters, the second column lists boundaries inside which the prior is uniform and nonzero (outside these boundaries, the prior probability is zero).  The third column lists median-likelihood values and credibility intervals that enclose the central $68\%$ ($95\%$) of area under the posterior PDF.  
\begin{table*}
\scriptsize
\centering
\caption{Summary of probability distribution functions for chemodynamical parameters}
\begin{tabular}{@{}llllllllll@{}}
\hline
parameter & prior & posterior & description\\
\hline
\smallskip
$\langle v_{\rm los}\rangle$ [km s$^{-1}$] & uniform between -500 and +500 & \vmeanexpanded & mean velocity at center\\
\smallskip
$\sigma_{v_{\rm los}}$ [km s$^{-1}$] & uniform between 0 and +500 & \vdispexpanded & velocity dispersion\\
\smallskip
$k_{v_{\rm los}}$ [km s$^{-1}$ arcmin$^{-1}$] & uniform between 0 and +10 & \vgradexpanded & magnitude of maximum velocity gradient\\
\smallskip
$\theta_{v_{\rm los}}$ [$\degr$] & uniform between -180 and +180 & \vthetaexpanded & direction of maximum velocity gradient\\
\smallskip
$\langle \feh\rangle $ & uniform between -5 and +1 & \fehmeanexpanded & mean metallicity at center\\
\smallskip
$\sigma_{\feh}$ & uniform between 0 and +2 & \fehdispexpanded & metallicity dispersion\\
\smallskip
$k_{\feh}$ [dex arcmin$^{-1}$] & uniform between -1 and +1 & \fehgradexpanded & magnitude of metallicity gradient\\
\hline
\end{tabular}
\\
\label{tab:gradient}
\end{table*}

\smallskip
We find that Ret 2 has mean line-of-sight velocity $\langle v_{\los}\rangle =$ \vmean\ km s$^{-1}$, with a resolved velocity dispersion of $\sigma_{v_{\los}}=$ \vdisp\ km s$^{-1}$.  We obtain a mean metallicity of $\langle \feh\rangle=$ \fehmean, with a dispersion of $\sigma_{\feh}=$ \fehdisp.  Taken at face value, the estimated metallicity dispersion is five times larger than is exhibited by any known globular cluster \citep{willman12}.  However, we note that the estimated metallicity dispersion is also similar to the median credibility interval for our metallicity estimates.  \citet{koposov11} demonstrate that, in such cases, estimates of intrinsic dispersion can be extremely sensitive to systematic over- or under-estimation of measurement errors.  Given the behavior in the bottom-right panel of Figure \ref{fig:ret2_repeats} (see Section \ref{subsec:repeatability} for discussion), we suspect that our metallicity errors are systematically overestimated, which would imply that our estimate of Ret 2's intrinsic metallicity dispersion is underestimated.  In any case, this estimate should be interpreted with caution and will need to be confirmed using spectra spanning a larger range of wavelengths.  

\smallskip  Finally, we do not detect significant gradients in either velocity or metallicity.  Our estimate of $k_{v_{\los}}=$ \vgrad km s$^{-1}$ arcmin$^{-1}$ is consistent with zero at the $\la 2\sigma$ level.  On the other hand, values as large as $\sim 1$ km s$^{-1}$ arcmin$^{-1}$ are similarly allowed.  The steepest allowed gradients are directed along position angle $\theta_{v_{\los}}\sim -80$ degrees, roughly $\sim 30$ degrees from Ret 2's morphological major axis (K15, DES15).  Our estimate of $k_{\feh}=$ \fehgrad dex arcmin$^{-1}$ is consistent with zero at the $\sim 1\sigma$ level, but also allows gradients as steep as $\pm 0.15$ dex arcmin$^{-1}$ within $\sim 2\sigma$.  We notice degeneracies between our estimates of dispersions and gradients for both velocity and metallicity distributions, such that larger gradients correspond to smaller dispersions.  However, our estimates for individual parameters (Table \ref{tab:gradient}) naturally include such effects, as the 1D PDFs for individual parameters are obtained by marginalizing over all other dimensions of the parameter space.

\subsection{Scaling Relations}
\label{subsec:scaling}

K15 use photometric data to show that Ret 2 occupies a region of structural parameter space that is populated by objects of ambiguous classification, somewhat intermediate between well-established globular clusters and dwarf galaxies \citep[][cf. K15's Figure 17]{gilmore07}.  Specifically, with projected halflight radius $R_{\rm h}=32\pm 1$ pc, Ret 2 is larger than nearly all globular clusters and smaller than nearly all known galaxies.  Moreover, its absolute magnitude, $M_V=-2.7\pm 0.1$, would place Ret 2 among the least luminous members of either population.  In these regards, Ret 2 and many of its newly-discovered siblings are similar to `ultrafaint' satellites Segue 1, Segue 2 and Willman 1, as well as to the extended globular clusters Pal 14 and Crater (\citealt{belokurov14,laevens14}, Mateo et al.\ 2015).  

\begin{figure}
  \includegraphics[width=5in,trim=0.in 0.1in 1.0in 0.5in,clip]{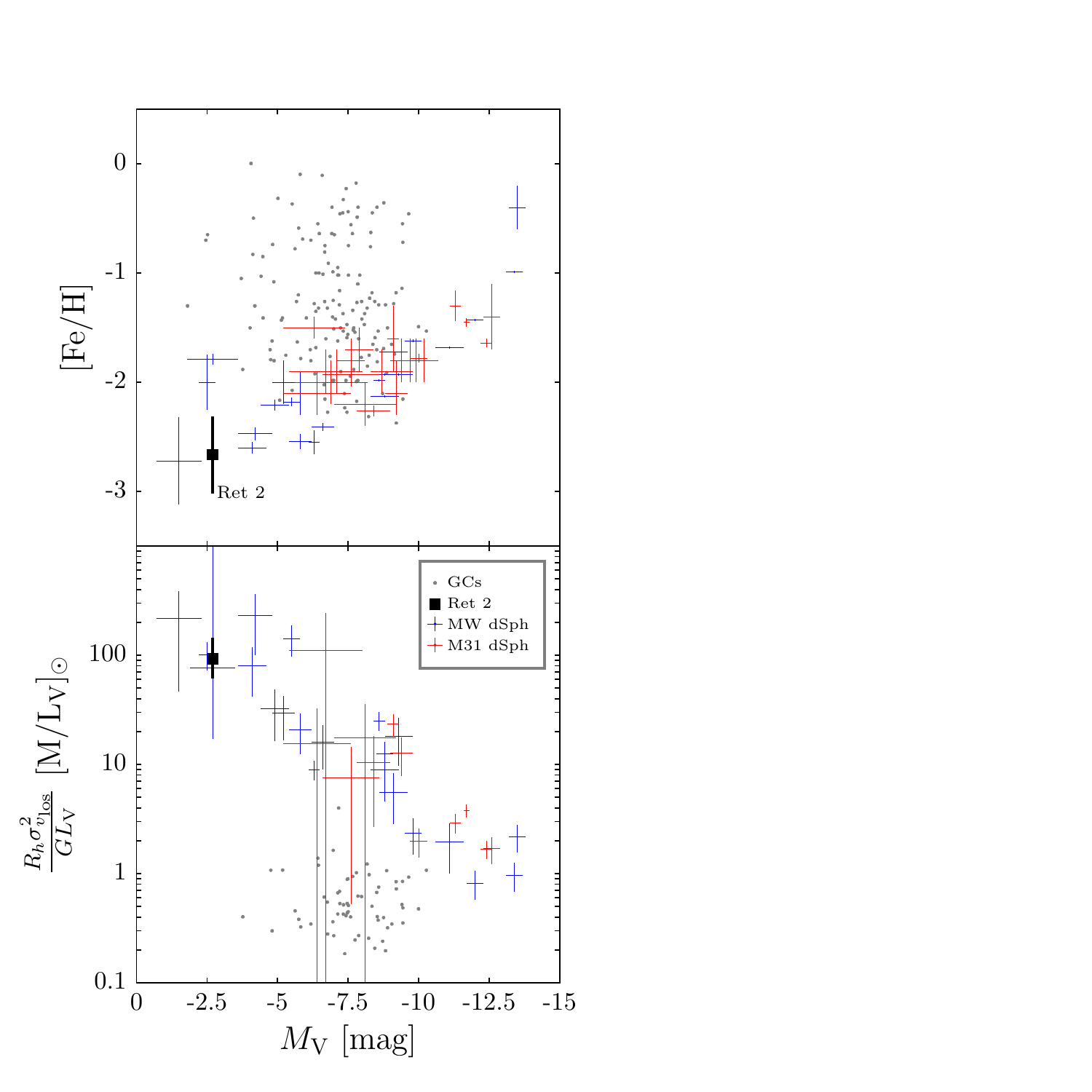}
  \caption{Mean metallicity and dynamical mass-to-light ratio vs absolute magnitude, for Galactic globular clusters (black points) as well as dwarf spheroidal satellites of the Milky Way (blue points with errorbars) and M31 (red points with errorbars).  Quantities plotted for Ret 2 are adopted from K15 and this work.  Data for globular clusters and dSphs are adopted, respectively, from the catalog of \citet[2010 edition]{harris96} and the review of \citet{mcconnachie12}.}
  \label{fig:ret2_scaling}
\end{figure}

\smallskip
Our spectroscopic results provide new leverage that can settle the question of Ret 2's nature.  For Galactic globular clusters as well as the dwarf spheroidal satellites of the Milky Way and M31, the top panel of Figure \ref{fig:ret2_scaling} plots mean metallicity against luminosity.  While globular clusters show no obvious trend, it is well-known that dwarf galaxies follow a luminosity/metallicity relation \citep{mateo98,tolstoy09,kirby13}.  Given the low mean metallicity that we estimate from the M2FS spectra, $\langle \feh\rangle=$ \fehmean, we place Ret 2 squarely onto the galactic relation (large black square in Figure \ref{fig:ret2_scaling}).

\smallskip
The lower panel of Figure \ref{fig:ret2_scaling} shows another well-known scaling relation that distinguishes dwarf galaxies from globular clusters.  Specifically, the mass-to-light ratios of dwarf galaxies are anti-correlated with luminosity, such that gravitational potentials in the least luminous galaxies all seem to be dominated by dark matter \citep{mateo93,walker07b,martin07,simon07}.  In contrast, the stellar kinematics of globular clusters generally do not require an internal dark matter component.  In order to see these trends, we plot $R_{\rm h}\sigma_{v_{\los}}^2/(GL_{\rm V})$ against luminosity, where 
$R_{\rm h}$ is halflight radius, $\sigma_{v_{\los}}$ is line-of-sight velocity dispersion, $L_{\rm V}$ is total V-band luminosity and $G$ is Newton's constant.  The combination of macroscopic observables on the vertical axis has dimensions of mass-to-light ratio\footnote{Many popular dynamical mass estimators have $M_{\rm dyn}\propto R_{\rm h}\sigma^2_{v_{\los}}/G$, where the constant of proportionality is typically between $\sim 2-6$ \citep[e.g., ][]{richstone86,wolf10,amorisco11b}.} and is therefore sufficient to highlight the different behavior of dwarf galaxies and globular clusters.  Again we find that Ret 2 follows the galactic relation.  Moreover, assuming dynamic equilibrium and negligible contamination from binary stars, Ret 2 has among the highest dynamical mass-to-light ratios of any known object.  The crude mass estimator of \citet{walker09d} implies that the dynamical mass enclosed within Ret 2's projected halflight radius is $M(R_{\rm h})\approx 5R_{\rm h}\sigma_{v_{\los}}^2/(2G)=$ \mrhalf\ $M_{\odot}$, and the associated mass-to-light ratio is $\approx 2M(R_{\rm h})/L_V=$ \mlratio\ in solar units.  

Table \ref{tab:summarytable} summarizes the observed properties of Ret 2, based on the photometric results of K15 and the spectroscopic results presented in this work.  Where photometric results from DES15 differ from those of K15 (e.g., for absolute magnitude), we list the DES15 results as well.
\begin{table*}
  \begin{centering}
  \scriptsize
  \caption{Summary of Reticulum 2's observed photometric and spectroscopic properties}
  \begin{tabular}{@{}lllllllll@{}}
    \hline
    Quantity&value&reference&notes\\
    \hline
    R.A. at center&$\alpha_{\rm J2000}=$03:35:42&K15$^{\dagger}$&\\
    Dec. at center&$\delta_{\rm J2000}=-$54:02:57&K15&\\
    Galactic longitude&$l=266.2958$ deg&K15\\
    Galactic latitude&$b=-49.7357$ deg&K15\\
    distance modulus&$m-M=17.4\pm 0.2$&K15\\
    distance from Sun&$D\sim 30$ kpc&K15\\
    absolute magnitude&$M_V=-2.7\pm 0.1$  ($-3.6\pm 0.1$)&K15 (DES15)\\
    exponential scale length& $R_{\rm e}=3.37_{-0.13}^{+0.23}$ arcmin&K15&semi-major axis\\
    ellipticity&$e=1-(b/a)=0.59^{+0.02}_{-0.03}$&K15\\
    position angle&PA$=71\pm 1$ deg&K15\\
    projected halflight radius&$R_{\rm h}=3.64_{-0.12}^{+0.21}$ arcmin &K15 & $R_{\rm h}\approx 1.68 R_e\sqrt{1-e}$\\
    projected halflight radius&$R_{\rm h}=32_{-1.1}^{+1.9}$ pc &K15\\
\\
    systemic line-of-sight velocity&$v_{\rm los}=$ \vmean\ km s$^{-1}$&this work& solar rest frame\\
    systemic line-of-sight velocity&$v_{\rm los}=-90.9$ km s$^{-1}$&this work& Galactic rest frame, given solar motion measured by \citet{schonrich10}\\
    internal velocity dispersion&$\sigma_{v_{\los}}=$ \vdisp\ km s$^{-1}$&this work\\
    velocity gradient & $k_{v_{\los}}=$ \vgrad\ km s$^{-1}$ arcmin$^{-1}$&this work\\
    PA of velocity gradient&$\theta_{v_{\los}}=$ \vtheta\ deg& this work\\
    mean metallicity&$\langle \feh\rangle=$ \fehmean\ dex&this work\\
    metallicity dispersion& $\sigma_{\feh}=$ \fehdisp\ dex&this work& similar to median metallicity error\\
    metallicity gradient&$k_{\feh}=$ \fehgrad\ dex arcmin$^{-1}$&this work\\
    mass enclosed within $R_{\rm h}$&$M(R_{\rm h})$= \mrhalf\ $M_{\odot}$&this work& $M(R_{\rm h})\approx 5R_{\rm h}\sigma_{v_{\los}}^2/(2G)$; assumes equilibrium, negligible binary stars\\
    mass-to-light raio&$\Upsilon=$ \mlratio\ $M_{\odot}/L_{\odot}$&this work&$\Upsilon\approx 2M(R_{\rm h})/L_V$; assumes equilibrium, negligible binary stars\\
    \hline
    $\dagger$ Unless otherwise noted, K15 and DES15 report similar values.
    \\
  \end{tabular}
  \label{tab:summarytable}
  \end{centering}
\end{table*}

\section{Summary and Discussion}
\label{sec:summary}

We have presented results from our initial stellar-spectroscopic observations of Ret 2.  Integrating for two hours in below-average observing conditions, Magellan/M2FS has delivered spectra sufficient for estimating $v_{\los}$, $\teff$, $\logg$ and $\feh$ for \goodnobs\ stars, reaching a limiting magnitude of $r\la 20.3$.  Based on a combination of photometrically- and spectroscopically-derived quantities, we have confirmed \nred\ likely members of Ret 2.  From the member sample we estimate a velocity dispersion of \vdisp\ km s$^{-1}$ about a mean of \vmean\ km s$^{-1}$ in the solar rest frame, and a metallicity dispersion of \fehdisp\ dex about a mean of \fehmean\ dex.  While Ret 2's size and luminosity are similar to those of globular clusters as well as ultrafaint dwarf galaxies, our estimates for mean metallicity and velocity dispersion place Ret 2 on scaling relations that are followed by dwarf galaxies and not by globular clusters.  On this basis, we conclude that Ret 2 is a bona fide galaxy.  

\smallskip
Furthermore, under assumptions of dynamic equilibrium and negligible contamination from binary stars, Ret 2's combination of size, velocity dispersion and luminosity implies a dynamical mass-to-light ratio of $\sim$ \mlratio $M_{\odot}/L_{\odot}$, among the largest inferred for any known object.  However, the validity of these assumptions remains unclear.  Perhaps undermining the assumption of dynamic equilibrium is Ret 2's significantly flattened morphology (K15, DES15), which---given Ret 2's Galactocentric distance of $\sim 30$ kpc---may signal ongoing tidal disruption.  On the other hand, our velocity sample does not show a significant velocity gradient that might reflect the ordered motions associated with tidal streaming \citep{pp95,read06}.  The mean velocity that we estimate for Ret 2 becomes $v_{\los}\sim -90.9$ km s$^{-1}$ in the Galactic rest frame, implying that Ret 2 is currently approaching pericenter.  We expect that deeper photometric and deeper/wider spectroscopic followup will be required in order to determine the extent to which Ret 2 is undergoing a disruptive tidal encounter with the Milky Way.

Finally, even if Ret 2 is resilient to tides, its velocity dispersion reflects some combination of its internal gravitational potential and the intrinsic velocity variability of binary stars.  \citet{mcconnachie10} have demonstrated that binary motions alone can contribute velocity dispersions of up to a few km s$^{-1}$.  Therefore we cannot exclude the possibility that Ret 2's velocity dispersion receives a significant contribution from binary stars, in which case the dynamical mass and mass-to-light ratios that we have reported would be systematically overestimated.  More epochs of spectroscopic observations of Ret 2---and other ultrafaint dwarfs with small velocity dispersions---are required in order to gauge the magnitude of this effect.  

\section*{Acknowledgements}

The construction of M2FS was supported by NSF/MRI grant AST-0923160, awarded to MM, Steve Shectman and Ian Thompson.  We thank Jeff Crane, Steve Shectman and Ian Thompson for invaluable assistance with the design, construction and support of M2FS.  We thank Josh Simon and the Dark Energy Survey's Milky Way Science Working Group for providing targeting coordinates for stars that they identified as probable members of Reticulum 2.  We also thank the DES collaboration for sharing a pre-submission version of their manuscript containing an independent analysis of the M2FS spectra for Ret 2; only after reading their paper did we decide to apply different zero-point corrections to velocities measured on blue and red channels.  MGW is supported by National Science Foundation grants AST-1313045, AST-1412999.  MM is supported by NSF grants AST-0808043 and AST-1312997.  EWO is supported by NSF grant AST-0807498 and AST-1313006.  The research leading to these results has received funding from the European Research Council under the European Union's Seventh Framework Programme (FP/2007-2013)/ERC Grant Agreement no. 308024.  This research has made use of the NASA/IPAC Extragalactic Database (NED), which is operated by the Jet Propulsion Laboratory, California Institute of Technology, under contract with the National Aeronautics and Space Administration.

\smallskip
Funding for the DES Projects has been provided by the U.S. Department of Energy, the U.S. National Science Foundation, the Ministry of Science and Education of Spain, 
the Science and Technology Facilities Council of the United Kingdom, the Higher Education Funding Council for England, the National Center for Supercomputing 
Applications at the University of Illinois at Urbana-Champaign, the Kavli Institute of Cosmological Physics at the University of Chicago, Financiadora de Estudos e Projetos, 
Funda{\c c}{\~a}o Carlos Chagas Filho de Amparo {\`a} Pesquisa do Estado do Rio de Janeiro, Conselho Nacional de Desenvolvimento Cient{\'i}fico e Tecnol{\'o}gico and 
the Minist{\'e}rio da Ci{\^e}ncia e Tecnologia, the Deutsche Forschungsgemeinschaft and the Collaborating Institutions in the Dark Energy Survey.
This work was supported in part by the U.S. Department of Energy contract to SLAC No.~DE-AC02-76SF00515.
The DES participants from Spanish institutions are partially supported by MINECO under grants AYA2012-39559, ESP2013-48274, FPA2013-47986, and Centro de Excelencia Severo Ochoa SEV-2012-0234, some of which include ERDF funds from the European Union.
ACR acknowledges financial support provided by the PAPDRJ CAPES/FAPERJ Fellowship.
AAP was supported by DOE grant DE-AC02-98CH10886 and by JPL, run by Caltech under a contract for NASA.

The Collaborating Institutions are Argonne National Laboratory, the University of California at Santa Cruz, the University of Cambridge, Centro de Investigaciones Energeticas, 
Medioambientales y Tecnologicas-Madrid, the University of Chicago, University College London, the DES-Brazil Consortium, the Eidgen{\"o}ssische Technische Hochschule (ETH) Z{\"u}rich, 
Fermi National Accelerator Laboratory, the University of Edinburgh, the University of Illinois at Urbana-Champaign, the Institut de Ciencies de l'Espai (IEEC/CSIC), 
the Institut de Fisica d'Altes Energies, Lawrence Berkeley National Laboratory, the Ludwig-Maximilians Universit{\"a}t and the associated Excellence Cluster Universe, 
the University of Michigan, the National Optical Astronomy Observatory, the University of Nottingham, The Ohio State University, the University of Pennsylvania, the University of Portsmouth, 
SLAC National Accelerator Laboratory, Stanford University, the University of Sussex, and Texas A\&M University.


\begin{thebibliography}{46}
\expandafter\ifx\csname natexlab\endcsname\relax\def\natexlab#1{#1}\fi

\bibitem[{{Amorisco} \& {Evans}(2011)}]{amorisco11b}
{Amorisco}, N.~C., \& {Evans}, N.~W. 2011, \mnras, 411, 2118

\bibitem[{{Belokurov} {et~al.}(2014){Belokurov}, {Irwin}, {Koposov}, {Evans},
  {Gonzalez-Solares}, {Metcalfe}, \& {Shanks}}]{belokurov14}
{Belokurov}, V., {Irwin}, M.~J., {Koposov}, S.~E., {Evans}, N.~W.,
  {Gonzalez-Solares}, E., {Metcalfe}, N., \& {Shanks}, T. 2014, \mnras, 441,
  2124

\bibitem[{{Belokurov et al.}(2007)}]{belokurov07}
{Belokurov et al.} 2007, \apj, 654, 897

\bibitem[{{Dotter} {et~al.}(2008){Dotter}, {Chaboyer}, {Jevremovi{\'c}},
  {Kostov}, {Baron}, \& {Ferguson}}]{dotter08}
{Dotter}, A., {Chaboyer}, B., {Jevremovi{\'c}}, D., {Kostov}, V., {Baron}, E.,
  \& {Ferguson}, J.~W. 2008, \apjs, 178, 89

\bibitem[{{Feroz} \& {Hobson}(2008)}]{feroz08}
{Feroz}, F., \& {Hobson}, M.~P. 2008, \mnras, 384, 449

\bibitem[{{Feroz} {et~al.}(2009){Feroz}, {Hobson}, \& {Bridges}}]{feroz09}
{Feroz}, F., {Hobson}, M.~P., \& {Bridges}, M. 2009, \mnras, 398, 1601

\bibitem[{{Geringer-Sameth} {et~al.}(2015){Geringer-Sameth}, {Walker},
  {Koushiappas}, {Koposov}, {Belokurov}, {Torrealba}, \& {Wyn Evans}}]{alex15}
{Geringer-Sameth}, A., {Walker}, M.~G., {Koushiappas}, S.~M., {Koposov}, S.~E.,
  {Belokurov}, V., {Torrealba}, G., \& {Wyn Evans}, N. 2015, ArXiv:1503.02320

\bibitem[{{Gilmore} {et~al.}(2007){Gilmore}, {Wilkinson}, {Wyse}, {Kleyna},
  {Koch}, {Evans}, \& {Grebel}}]{gilmore07}
{Gilmore}, G., {Wilkinson}, M.~I., {Wyse}, R.~F.~G., {Kleyna}, J.~T., {Koch},
  A., {Evans}, N.~W., \& {Grebel}, E.~K. 2007, \apj, 663, 948

\bibitem[{{Harris}(1996)}]{harris96}
{Harris}, W.~E. 1996, \aj, 112, 1487

\bibitem[{{Hooper} \& {Linden}(2015)}]{hooper15}
{Hooper}, D., \& {Linden}, T. 2015, ArXiv:1503.06209

\bibitem[{{Kirby} {et~al.}(2013){Kirby}, {Cohen}, {Guhathakurta}, {Cheng},
  {Bullock}, \& {Gallazzi}}]{kirby13}
{Kirby}, E.~N., {Cohen}, J.~G., {Guhathakurta}, P., {Cheng}, L., {Bullock},
  J.~S., \& {Gallazzi}, A. 2013, \apj, 779, 102

\bibitem[{{Koleva} {et~al.}(2009){Koleva}, {Prugniel}, {Bouchard}, \&
  {Wu}}]{koleva09}
{Koleva}, M., {Prugniel}, P., {Bouchard}, A., \& {Wu}, Y. 2009, \aap, 501, 1269

\bibitem[{{Koposov} {et~al.}(2015){Koposov}, {Belokurov}, {Torrealba}, \& {Wyn
  Evans}}]{koposov15}
{Koposov}, S.~E., {Belokurov}, V., {Torrealba}, G., \& {Wyn Evans}, N. 2015,
  ArXiv:1503.02079

\bibitem[{{Koposov} {et~al.}(2011){Koposov}, {Gilmore}, {Walker}, {Belokurov},
  {Wyn Evans}, {Fellhauer}, {Gieren}, {Geisler}, {Monaco}, {Norris}, {Okamoto},
  {Pe{\~n}arrubia}, {Wilkinson}, {Wyse}, \& {Zucker}}]{koposov11}
{Koposov}, S.~E., {Gilmore}, G., {Walker}, M.~G., {Belokurov}, V., {Wyn Evans},
  N., {Fellhauer}, M., {Gieren}, W., {Geisler}, D., {Monaco}, L., {Norris},
  J.~E., {Okamoto}, S., {Pe{\~n}arrubia}, J., {Wilkinson}, M., {Wyse},
  R.~F.~G., \& {Zucker}, D.~B. 2011, \apj, 736, 146

\bibitem[{{Laevens} {et~al.}(2014){Laevens}, {Martin}, {Sesar}, {Bernard},
  {Rix}, {Slater}, {Bell}, {Ferguson}, {Schlafly}, {Burgett}, {Chambers},
  {Denneau}, {Draper}, {Kaiser}, {Kudritzki}, {Magnier}, {Metcalfe}, {Morgan},
  {Price}, {Sweeney}, {Tonry}, {Wainscoat}, \& {Waters}}]{laevens14}
{Laevens}, B.~P.~M., {Martin}, N.~F., {Sesar}, B., {Bernard}, E.~J., {Rix},
  H.-W., {Slater}, C.~T., {Bell}, E.~F., {Ferguson}, A.~M.~N., {Schlafly},
  E.~F., {Burgett}, W.~S., {Chambers}, K.~C., {Denneau}, L., {Draper}, P.~W.,
  {Kaiser}, N., {Kudritzki}, R.-P., {Magnier}, E.~A., {Metcalfe}, N., {Morgan},
  J.~S., {Price}, P.~A., {Sweeney}, W.~E., {Tonry}, J.~L., {Wainscoat}, R.~J.,
  \& {Waters}, C. 2014, \apjl, 786, L3

\bibitem[{{Lee} {et~al.}(2008{\natexlab{a}}){Lee}, {Beers}, {Sivarani},
  {Allende Prieto}, {Koesterke}, {Wilhelm}, {Re Fiorentin}, {Bailer-Jones},
  {Norris}, {Rockosi}, {Yanny}, {Newberg}, {Covey}, {Zhang}, \& {Luo}}]{lee08a}
{Lee}, Y.~S., {Beers}, T.~C., {Sivarani}, T., {Allende Prieto}, C.,
  {Koesterke}, L., {Wilhelm}, R., {Re Fiorentin}, P., {Bailer-Jones}, C.~A.~L.,
  {Norris}, J.~E., {Rockosi}, C.~M., {Yanny}, B., {Newberg}, H.~J., {Covey},
  K.~R., {Zhang}, H.-T., \& {Luo}, A.-L. 2008{\natexlab{a}}, \aj, 136, 2022

\bibitem[{{Lee} {et~al.}(2008{\natexlab{b}}){Lee}, {Beers}, {Sivarani},
  {Johnson}, {An}, {Wilhelm}, {Allende Prieto}, {Koesterke}, {Re Fiorentin},
  {Bailer-Jones}, {Norris}, {Yanny}, {Rockosi}, {Newberg}, {Cudworth}, \&
  {Pan}}]{lee08b}
{Lee}, Y.~S., {Beers}, T.~C., {Sivarani}, T., {Johnson}, J.~A., {An}, D.,
  {Wilhelm}, R., {Allende Prieto}, C., {Koesterke}, L., {Re Fiorentin}, P.,
  {Bailer-Jones}, C.~A.~L., {Norris}, J.~E., {Yanny}, B., {Rockosi}, C.,
  {Newberg}, H.~J., {Cudworth}, K.~M., \& {Pan}, K. 2008{\natexlab{b}}, \aj,
  136, 2050

\bibitem[{{Martin} {et~al.}(2008){Martin}, {de Jong}, \& {Rix}}]{martin08}
{Martin}, N.~F., {de Jong}, J.~T.~A., \& {Rix}, H.-W. 2008, \apj, 684, 1075

\bibitem[{{Martin} {et~al.}(2007){Martin}, {Ibata}, {Chapman}, {Irwin}, \&
  {Lewis}}]{martin07}
{Martin}, N.~F., {Ibata}, R.~A., {Chapman}, S.~C., {Irwin}, M., \& {Lewis},
  G.~F. 2007, \mnras, 380, 281

\bibitem[{{Martin} {et~al.}(2013){Martin}, {Ibata}, {McConnachie}, {Mackey},
  {Ferguson}, {Irwin}, {Lewis}, \& {Fardal}}]{martin13}
{Martin}, N.~F., {Ibata}, R.~A., {McConnachie}, A.~W., {Mackey}, A.~D.,
  {Ferguson}, A.~M.~N., {Irwin}, M.~J., {Lewis}, G.~F., \& {Fardal}, M.~A.
  2013, \apj, 776, 80

\bibitem[{{Mateo} {et~al.}(2012){Mateo}, {Bailey}, {Crane}, {Shectman},
  {Thompson}, {Roederer}, {Bigelow}, \& {Gunnels}}]{mateo12}
{Mateo}, M., {Bailey}, J.~I., {Crane}, J., {Shectman}, S., {Thompson}, I.,
  {Roederer}, I., {Bigelow}, B., \& {Gunnels}, S. 2012, in Society of
  Photo-Optical Instrumentation Engineers (SPIE) Conference Series, Vol. 8446,
  Society of Photo-Optical Instrumentation Engineers (SPIE) Conference Series,
  4

\bibitem[{{Mateo} {et~al.}(1993){Mateo}, {Olszewski}, {Pryor}, {Welch}, \&
  {Fischer}}]{mateo93}
{Mateo}, M., {Olszewski}, E.~W., {Pryor}, C., {Welch}, D.~L., \& {Fischer}, P.
  1993, \aj, 105, 510

\bibitem[{{Mateo}(1998)}]{mateo98}
{Mateo}, M.~L. 1998, \araa, 36, 435

\bibitem[{{McConnachie}(2012)}]{mcconnachie12}
{McConnachie}, A.~W. 2012, \aj, 144, 4

\bibitem[{{McConnachie} \& {C{\^o}t{\'e}}(2010)}]{mcconnachie10}
{McConnachie}, A.~W., \& {C{\^o}t{\'e}}, P. 2010, \apjl, 722, L209

\bibitem[{{McConnachie} {et~al.}(2009){McConnachie}, {Irwin}, {Ibata},
  {Dubinski}, {Widrow}, {Martin}, {C{\^o}t{\'e}}, {Dotter}, {Navarro},
  {Ferguson}, {Puzia}, {Lewis}, {Babul}, {Barmby}, {Bienaym{\'e}}, {Chapman},
  {Cockcroft}, {Collins}, {Fardal}, {Harris}, {Huxor}, {Mackey},
  {Pe{\~n}arrubia}, {Rich}, {Richer}, {Siebert}, {Tanvir}, {Valls-Gabaud}, \&
  {Venn}}]{mcconnachie09}
{McConnachie}, A.~W., {Irwin}, M.~J., {Ibata}, R.~A., {Dubinski}, J., {Widrow},
  L.~M., {Martin}, N.~F., {C{\^o}t{\'e}}, P., {Dotter}, A.~L., {Navarro},
  J.~F., {Ferguson}, A.~M.~N., {Puzia}, T.~H., {Lewis}, G.~F., {Babul}, A.,
  {Barmby}, P., {Bienaym{\'e}}, O., {Chapman}, S.~C., {Cockcroft}, R.,
  {Collins}, M.~L.~M., {Fardal}, M.~A., {Harris}, W.~E., {Huxor}, A., {Mackey},
  A.~D., {Pe{\~n}arrubia}, J., {Rich}, R.~M., {Richer}, H.~B., {Siebert}, A.,
  {Tanvir}, N., {Valls-Gabaud}, D., \& {Venn}, K.~A. 2009, \nat, 461, 66

\bibitem[{{Piatek} \& {Pryor}(1995)}]{pp95}
{Piatek}, S., \& {Pryor}, C. 1995, \aj, 109, 1071

\bibitem[{{Read} {et~al.}(2006){Read}, {Wilkinson}, {Evans}, {Gilmore}, \&
  {Kleyna}}]{read06}
{Read}, J.~I., {Wilkinson}, M.~I., {Evans}, N.~W., {Gilmore}, G., \& {Kleyna},
  J.~T. 2006, \mnras, 366, 429

\bibitem[{{Richstone} \& {Tremaine}(1986)}]{richstone86}
{Richstone}, D.~O., \& {Tremaine}, S. 1986, \aj, 92, 72

\bibitem[{{Schlegel} {et~al.}(1998){Schlegel}, {Finkbeiner}, \&
  {Davis}}]{schlegel98}
{Schlegel}, D.~J., {Finkbeiner}, D.~P., \& {Davis}, M. 1998, \apj, 500, 525

\bibitem[{{Sch{\"o}nrich} {et~al.}(2010){Sch{\"o}nrich}, {Binney}, \&
  {Dehnen}}]{schonrich10}
{Sch{\"o}nrich}, R., {Binney}, J., \& {Dehnen}, W. 2010, \mnras, 403, 1829

\bibitem[{{Simon} \& {Geha}(2007)}]{simon07}
{Simon}, J.~D., \& {Geha}, M. 2007, \apj, 670, 313

\bibitem[{{Srianand} {et~al.}(2010){Srianand}, {Gupta}, {Petitjean},
  {Noterdaeme}, \& {Ledoux}}]{srianand10}
{Srianand}, R., {Gupta}, N., {Petitjean}, P., {Noterdaeme}, P., \& {Ledoux}, C.
  2010, \mnras, 405, 1888

\bibitem[{{The DES Collaboration} {et~al.}(2015){The DES Collaboration},
  {Bechtol}, {Drlica-Wagner}, {Balbinot}, {Pieres}, {Simon}, {Yanny},
  {Santiago}, {Wechsler}, {Frieman}, {Walker}, {Williams}, {Rozo}, {Rykoff},
  {Queiroz}, {Luque}, {Benoit-Levy}, {Bernstein}, {Tucker}, {Sevilla},
  {Gruendl}, {da Costa}, {Fausti Neto}, {Maia}, {Abbott}, {Allam}, {Armstrong},
  {Bauer}, {Bernstein}, {Bertin}, {Brooks}, {Buckley-Geer}, {Burke}, {Carnero
  Rosell}, {Castander}, {D'Andrea}, {DePoy}, {Desai}, {Diehl}, {Eifler},
  {Estrada}, {Evrard}, {Fernandez}, {Finley}, {Flaugher}, {Gaztanaga},
  {Gerdes}, {Girardi}, {Gladders}, {Gruen}, {Gutierrez}, {Hao}, {Honscheid},
  {Jain}, {James}, {Kent}, {Kron}, {Kuehn}, {Kuropatkin}, {Lahav}, {Li}, {Lin},
  {Makler}, {March}, {Marshall}, {Martini}, {Merritt}, {Miller}, {Miquel},
  {Mohr}, {Neilsen}, {Nichol}, {Nord}, {Ogando}, {Peoples}, {Petravick},
  {Plazas}, {Romer}, {Roodman}, {Sako}, {Sanchez}, {Scarpine}, {Schubnell},
  {Smith}, {Soares-Santos}, {Sobreira}, {Suchyta}, {Swanson}, {Tarle},
  {Thaler}, {Thomas}, {Wester}, \& {Zuntz}}]{des15}
{The DES Collaboration}, {Bechtol}, K., {Drlica-Wagner}, A., {Balbinot}, E.,
  {Pieres}, A., {Simon}, J.~D., {Yanny}, B., {Santiago}, B., {Wechsler}, R.~H.,
  {Frieman}, J., {Walker}, A.~R., {Williams}, P., {Rozo}, E., {Rykoff}, E.~S.,
  {Queiroz}, A., {Luque}, E., {Benoit-Levy}, A., {Bernstein}, R.~A., {Tucker},
  D., {Sevilla}, I., {Gruendl}, R.~A., {da Costa}, L.~N., {Fausti Neto}, A.,
  {Maia}, M.~A.~G., {Abbott}, T., {Allam}, S., {Armstrong}, R., {Bauer}, A.~H.,
  {Bernstein}, G.~M., {Bertin}, E., {Brooks}, D., {Buckley-Geer}, E., {Burke},
  D.~L., {Carnero Rosell}, A., {Castander}, F.~J., {D'Andrea}, C.~B., {DePoy},
  D.~L., {Desai}, S., {Diehl}, H.~T., {Eifler}, T.~F., {Estrada}, J., {Evrard},
  A.~E., {Fernandez}, E., {Finley}, D.~A., {Flaugher}, B., {Gaztanaga}, E.,
  {Gerdes}, D., {Girardi}, L., {Gladders}, M., {Gruen}, D., {Gutierrez}, G.,
  {Hao}, J., {Honscheid}, K., {Jain}, B., {James}, D., {Kent}, S., {Kron}, R.,
  {Kuehn}, K., {Kuropatkin}, N., {Lahav}, O., {Li}, T.~S., {Lin}, H., {Makler},
  M., {March}, M., {Marshall}, J., {Martini}, P., {Merritt}, K.~W., {Miller},
  C., {Miquel}, R., {Mohr}, J., {Neilsen}, E., {Nichol}, R., {Nord}, B.,
  {Ogando}, R., {Peoples}, J., {Petravick}, D., {Plazas}, A.~A., {Romer},
  A.~K., {Roodman}, A., {Sako}, M., {Sanchez}, E., {Scarpine}, V., {Schubnell},
  M., {Smith}, R.~C., {Soares-Santos}, M., {Sobreira}, F., {Suchyta}, E.,
  {Swanson}, M.~E.~C., {Tarle}, G., {Thaler}, J., {Thomas}, D., {Wester}, W.,
  \& {Zuntz}, J. 2015, ArXiv:1503.02584

\bibitem[{{The Fermi-LAT Collaboration} {et~al.}(2015){The Fermi-LAT
  Collaboration}, {The DES Collaboration}, {:}, {Drlica-Wagner}, {Albert},
  {Bechtol}, {Wood}, {Strigari}, {Sanchez-Conde}, {Baldini}, {Essig},
  {Cohen-Tanugi}, {Anderson}, {Bellazzini}, {Bloom}, {Caputo}, {Cecchi},
  {Charles}, {Chiang}, {Conrad}, {de Angelis}, {Funk}, {Fusco}, {Gargano},
  {Giglietto}, {Giordano}, {Guiriec}, {Gustafsson}, {Kuss}, {Loparco},
  {Lubrano}, {Mirabal}, {Mizuno}, {Morselli}, {Ohsugi}, {Orlando}, {Persic},
  {Raino}, {Spada}, {Suson}, {Zaharijas}, {Zimmer}, {Abbott}, {Allam},
  {Balbinot}, {Bauer}, {Benoit-Levy}, {Bernstein}, {Bernstein}, {Bertin},
  {Brooks}, {Buckley-Geer}, {Burke}, {Carnero Rosell}, {Castander},
  {Covarrubias}, {D'Andrea}, {da Costa}, {DePoy}, {Desai}, {Diehl}, {Cunha},
  {Eifler}, {Estrada}, {Evrard}, {Fausti Neto}, {Fernandez}, {Finley},
  {Flaugher}, {Frieman}, {Gaztanaga}, {Gerdes}, {Gruen}, {Gruendl},
  {Gutierrez}, {Honscheid}, {Jain}, {James}, {Jeltema}, {Kent}, {Kron},
  {Kuropatkin}, {Lahav}, {Li}, {Luque}, {Maia}, {Makler}, {March}, {Marshall},
  {Martini}, {Merritt}, {Miller}, {Miquel}, {Mohr}, {Neilsen}, {Nord},
  {Ogando}, {Peoples}, {Petravick}, {Pieres}, {Plazas}, {Queiroz}, {Romer},
  {Roodman}, {Rykoff}, {Sako}, {Sanchez}, {Santiago}, {Scarpine}, {Schubnell},
  {Sevilla}, {Smith}, {Soares-Santos}, {Sobreira}, {Suchyta}, {Swanson},
  {Tarle}, {Thaler}, {Thomas}, {Tucker}, {Walker}, {Wechsler}, {Wester},
  {Williams}, {Yanny}, \& {Zuntz}}]{fermi15}
{The Fermi-LAT Collaboration}, {The DES Collaboration}, {:}, {Drlica-Wagner},
  A., {Albert}, A., {Bechtol}, K., {Wood}, M., {Strigari}, L., {Sanchez-Conde},
  M., {Baldini}, L., {Essig}, R., {Cohen-Tanugi}, J., {Anderson}, B.,
  {Bellazzini}, R., {Bloom}, E.~D., {Caputo}, R., {Cecchi}, C., {Charles}, E.,
  {Chiang}, J., {Conrad}, J., {de Angelis}, A., {Funk}, S., {Fusco}, P.,
  {Gargano}, F., {Giglietto}, N., {Giordano}, F., {Guiriec}, S., {Gustafsson},
  M., {Kuss}, M., {Loparco}, F., {Lubrano}, P., {Mirabal}, N., {Mizuno}, T.,
  {Morselli}, A., {Ohsugi}, T., {Orlando}, E., {Persic}, M., {Raino}, S.,
  {Spada}, F., {Suson}, D.~J., {Zaharijas}, G., {Zimmer}, S., {Abbott}, T.,
  {Allam}, S., {Balbinot}, E., {Bauer}, A.~H., {Benoit-Levy}, A., {Bernstein},
  R.~A., {Bernstein}, G.~M., {Bertin}, E., {Brooks}, D., {Buckley-Geer}, E.,
  {Burke}, D.~L., {Carnero Rosell}, A., {Castander}, F.~J., {Covarrubias}, R.,
  {D'Andrea}, C.~B., {da Costa}, L.~N., {DePoy}, D.~L., {Desai}, S., {Diehl},
  H.~T., {Cunha}, C.~E., {Eifler}, T.~F., {Estrada}, J., {Evrard}, A.~E.,
  {Fausti Neto}, A., {Fernandez}, E., {Finley}, D.~A., {Flaugher}, B.,
  {Frieman}, J., {Gaztanaga}, E., {Gerdes}, D., {Gruen}, D., {Gruendl}, R.~A.,
  {Gutierrez}, G., {Honscheid}, K., {Jain}, B., {James}, D., {Jeltema}, T.,
  {Kent}, S., {Kron}, R., {Kuropatkin}, N., {Lahav}, O., {Li}, T.~S., {Luque},
  E., {Maia}, M.~A.~G., {Makler}, M., {March}, M., {Marshall}, J., {Martini},
  P., {Merritt}, K.~W., {Miller}, C., {Miquel}, R., {Mohr}, J., {Neilsen}, E.,
  {Nord}, B., {Ogando}, R., {Peoples}, J., {Petravick}, D., {Pieres}, A.,
  {Plazas}, A.~A., {Queiroz}, A., {Romer}, A.~K., {Roodman}, A., {Rykoff},
  E.~S., {Sako}, M., {Sanchez}, E., {Santiago}, B., {Scarpine}, V.,
  {Schubnell}, M., {Sevilla}, I., {Smith}, R.~C., {Soares-Santos}, M.,
  {Sobreira}, F., {Suchyta}, E., {Swanson}, M.~E.~C., {Tarle}, G., {Thaler},
  J., {Thomas}, D., {Tucker}, D., {Walker}, A., {Wechsler}, R.~H., {Wester},
  W., {Williams}, P., {Yanny}, B., \& {Zuntz}, J. 2015, ArXiv:1503.02632

\bibitem[{{Tolstoy} {et~al.}(2009){Tolstoy}, {Hill}, \& {Tosi}}]{tolstoy09}
{Tolstoy}, E., {Hill}, V., \& {Tosi}, M. 2009, \araa, 47, 371

\bibitem[{{Udry} {et~al.}(1999){Udry}, {Mayor}, {Maurice}, {Andersen},
  {Imbert}, {Lindgren}, {Mermilliod}, {Nordstr{\"o}m}, \&
  {Pr{\'e}vot}}]{udry99}
{Udry}, S., {Mayor}, M., {Maurice}, E., {Andersen}, J., {Imbert}, M.,
  {Lindgren}, H., {Mermilliod}, J.-C., {Nordstr{\"o}m}, B., \& {Pr{\'e}vot}, L.
  1999, in ASP Conf. Ser. 185: IAU Colloq. 170: Precise Stellar Radial
  Velocities, ed. J.~B. {Hearnshaw} \& C.~D. {Scarfe}, 383--+

\bibitem[{{Walker} {et~al.}(2007){Walker}, {Mateo}, {Olszewski}, {Gnedin},
  {Wang}, {Sen}, \& {Woodroofe}}]{walker07b}
{Walker}, M.~G., {Mateo}, M., {Olszewski}, E.~W., {Gnedin}, O.~Y., {Wang}, X.,
  {Sen}, B., \& {Woodroofe}, M. 2007, \apjl, 667, L53

\bibitem[{{Walker} {et~al.}(2009{\natexlab{a}}){Walker}, {Mateo}, {Olszewski},
  {Pe{\~n}arrubia}, {Wyn Evans}, \& {Gilmore}}]{walker09d}
{Walker}, M.~G., {Mateo}, M., {Olszewski}, E.~W., {Pe{\~n}arrubia}, J., {Wyn
  Evans}, N., \& {Gilmore}, G. 2009{\natexlab{a}}, \apj, 704, 1274

\bibitem[{{Walker} {et~al.}(2009{\natexlab{b}}){Walker}, {Mateo}, {Olszewski},
  {Sen}, \& {Woodroofe}}]{walker09b}
{Walker}, M.~G., {Mateo}, M., {Olszewski}, E.~W., {Sen}, B., \& {Woodroofe}, M.
  2009{\natexlab{b}}, \aj, 137, 3109

\bibitem[{{Walker} {et~al.}(2015){Walker}, {Olszewski}, \& {Mateo}}]{walker15}
{Walker}, M.~G., {Olszewski}, E.~W., \& {Mateo}, M. 2015, ArXiv:1503.02589

\bibitem[{{Weck} {et~al.}(2003){Weck}, {Schweitzer}, {Stancil}, {Hauschildt},
  \& {Kirby}}]{weck03}
{Weck}, P.~F., {Schweitzer}, A., {Stancil}, P.~C., {Hauschildt}, P.~H., \&
  {Kirby}, K. 2003, \apj, 582, 1059

\bibitem[{{Willman} \& {Strader}(2012)}]{willman12}
{Willman}, B., \& {Strader}, J. 2012, \aj, 144, 76

\bibitem[{{Willman et al.}(2005)}]{willman05b}
{Willman et al.} 2005, \aj, 129, 2692

\bibitem[{{Wolf} {et~al.}(2010){Wolf}, {Martinez}, {Bullock}, {Kaplinghat},
  {Geha}, {Mu{\~n}oz}, {Simon}, \& {Avedo}}]{wolf10}
{Wolf}, J., {Martinez}, G.~D., {Bullock}, J.~S., {Kaplinghat}, M., {Geha}, M.,
  {Mu{\~n}oz}, R.~R., {Simon}, J.~D., \& {Avedo}, F.~F. 2010, \mnras, 406, 1220

\bibitem[{{Zucker et al.}(2006)}]{zucker06a}
{Zucker et al.} 2006, \apjl, 643, L103

\end{thebibliography}

\end{document}